\documentclass[jcp,aps,preprint]{revtex4}
\usepackage{xcolor}
\usepackage{amsmath}
\usepackage{amstext,bm}
\usepackage{epsfig}
\usepackage{xcolor}
\usepackage{multirow}
\usepackage{rotating}
\definecolor{dgreen}{rgb}{0,.5,0}
\definecolor{dred}{rgb}{.7,.0,.0}
\def\etal{\mbox{et al.}}
\newcommand{\manu}[1]{{\textcolor{dgreen}{ Manu: #1 }} }

\newcommand{\supi}[2]{#1^{\rm #2}}
\newcommand{\subsupi}[3]{#1_{\rm #2}^{\rm #3}}

\begin{document}
\markboth{Yann Cornaton {\it et al.}}{Molecular Physics}

\title{{\itshape 
Analysis of double hybrid density-functionals along the adiabatic
connection
}
}

\author{
Yann Cornaton$^{a}$, Odile Franck$^{a}$, Andrew M.~Teale$^{b, c}$ and
Emmanuel Fromager$^{a}$$^{\ast}$\thanks{$^\ast$Corresponding author.
Email: fromagere@unistra.fr 
\vspace{6pt}}
\\\vspace{6pt}  
$^{a}${\em{
Laboratoire de Chimie Quantique,
Institut de Chimie, CNRS / Universit\'{e} de Strasbourg,
4 rue Blaise Pascal, 67000 Strasbourg, France
}};\\\vspace{6pt}  
$^{b}${\em{
School of Chemistry, University of Nottingham, University Park, Nottingham, NG7 2RD,
 United Kingdom
}};\\\vspace{6pt}  
$^{c}${\em{
Department of Chemistry, Centre for Theoretical and Computational Chemistry,
University of Oslo, P.~O.~Box 1033, Blindern, Oslo N-0315, Norway
}}
\\
}

\begin{abstract}

We present a graphical analysis of the adiabatic connections underlying double-hybrid density-functional methods that employ second-order perturbation theory. Approximate adiabatic connection formulae relevant to the construction of these functionals are derived and compared directly with those calculated using accurate \textit{ab initio} methods. The discontinuous nature of the approximate adiabatic integrands is emphasized, the discontinuities occurring at interaction strengths which mark the transitions between regions that are: (i) described predominantly by second-order perturbation theory (ii) described by a mixture of density-functional and second-order perturbation theory contributions and (iii) described purely by density-functional theory. Numerical examples are presented for a selection of small molecular systems and van der Waals dimers. The impacts of commonly used approximations in each of the three sections of the adiabatic connection are discussed along with possible routes for the development of improved double-hybrid methodologies.
\bigskip

\bigskip

\end{abstract}

\maketitle

\section{Introduction}\label{sec:intro}

Hybrid density-functional methodologies are now amongst the most commonly applied methods in quantum chemistry. The idea to hybridize density-functional theory with wave-function based approaches can be traced back to the work of Becke~\cite{BECKE:1993p1811,BECKE:1993p1788}. Motivated by the adiabatic connection (AC) 
formalism~\cite{LANGRETH:1975p1425,GUNNARSSON:1976p1781,Gunnarsson:1977,LANGRETH:1977p1780,SAVIN:2003p635}
Becke suggested that the introduction of a fraction of orbital dependent
exchange could be beneficial and deliver improved performance over
standard generalized gradient approximation (GGA) functionals. Becke
initially constructed a half-and-half functional containing 50\% orbital dependent exchange and 50\% of a density-functional approximation to exchange based on a simple linear model of the AC. Although the performance of this initial model was disappointing subsequent empirical optimization of the weight of orbital dependent exchange based on thermochemical performance showed that a weight in the range 20\%--30\% delivered improved performance over standard methods. Later, a number of theoretical rationales for weights of orbital dependent exchange in this range have been put forward~\cite{PERDEW:1996p1804,Burke:1997p643}, however, as recently noted by Cortona~\cite{Cortona:2012} and utilized by Guido \emph{et. al}~\cite{Guido:2013}, similar arguments can be made for a number of different values.  

The widespread adoption of exchange-based hybrid functionals for molecular applications can be understood from their improved performance in applications such as thermochemistry and the determination of equilibrium molecular structures. Perhaps more crucially, the extra computational effort associated with the evaluation of the orbital dependent exchange is sufficiently modest so as not to hinder the application of hybrid approaches to a wide variety of chemical problems. Nonetheless, the introduction of orbital dependent exchange is not a panacea for all issues associated with standard GGA functionals. For many properties, or for molecules at geometries away from their equilibrium structure, the inclusion of orbital dependent exchange may be detrimental. For the simple example of the H$_2$ molecule see Ref.~\cite{GRITSENKO:1996p1796}. Furthermore, it is well known that error cancellations play a crucial role when combining GGA exchange and correlation components~\cite{Baerends:1997p579} and the combination of 100\% orbital dependent exchange with typical GGA correlation functionals is therefore ineffective. The construction of optimal hybrid exchange-based methods requires a tradeoff between improving the description of the exchange component whilst maintaining a reasonable description of correlation effects.

A natural extension of hybrid approaches is to consider the hybridization
of the correlation energy in addition to exchange energy component,
resulting in so-called double hybrids. In recent years a number of
empirical approaches which consider the post SCF addition of
second-order perturbation theory (PT2) terms have been
developed~\cite{2blehybrids_Grimme,mPW2_Grimme,b2t_Tarno,B2pi_Garcia,pbe0-dh_Adamo,B2GP_Martin,Xin_XYG3_2009,Xin_XYGJ-OS_2011,Xin_Adamo_xDH-PBE0},
perhaps the most well known of which is the B2-PLYP functional of
Grimme~\cite{2blehybrids_Grimme}. More rigorous routes to combine DFT
and PT2 approaches have followed the suggestion by Savin~\cite{SavBook}
to employ long-range only wave function approaches in tandem with
specially designed short-range density functionals. From a
density-functional point of view this approach can be justified by use
of a generalized AC as shown by
Yang~\cite{Yang:1998p441}. In this manner implementations have been
constructed which combine most of the models of quantum chemistry with
density-functional theory including,  configuration-interaction
theory~\cite{Leininger:1997p2197,Pollet:2002p2223}, M{\o}ller--Plesset
(MP) perturbation theory~\cite{Angyan:2005p2198}, coupled-cluster theory~\cite{Goll:2005p2199}, multi-configurational self-consistent field theory~\cite{Fromager:2007p1946,Fromager:2009p2200} and N-electron valence state perturbation theory~\cite{nevpt2srdft}. 

The advantage of range-separated approaches is that the division of
labour between the wave function and density-functional methods can to
some extent be controlled by the manner in which the two-electron
integrals are modified, typically by using an error-function
attenuation~\cite{Pollet:2002p1800,Toulouse:2005p1774}. This has been
shown, for example, to be effective in the treatment of dispersion
interactions~\cite{Angyan:2005p2198,Teale:2011}. The disadvantage of
such an approach is the need to develop specialized complementary
density functionals, although this task has been undertaken at the
local density approximation (LDA)~\cite{SavBook,Toulouse:2004p1959}, GGA~\cite{Goll:2005p2199,Toulouse:2005p2201,Goll:2006p2202} and meta-GGA~\cite{Goll:2009p2219} levels. 

Recently, Sharkas \emph{et al.}~\cite{2blehybrids_Julien} have used a
similar approach to consider the standard linear AC to
hybridize DFT with PT2 for functionals in which the fraction $a_{\rm c}$ of MP2
correlation energy equals the square of the fraction $a_{\rm x}$ of
Hartree--Fock (HF) exchange. Fromager~\cite{2blehybrids_Fromager2011JCP} has recently extended this approach to incorporate more flexible two-parameter double-hybrid
energy expressions that satisfy the inequality $a_{\rm c}\leq a^2_{\rm x}$. Commonly used empirically optimized double-hybrid methods satisfy this inequality but do not in general have values of $a_{\rm c}$ exactly equal to the $a^2_{\rm x}$. This approach allows the construction and analysis of energy expressions, similar to those used in empirical double-hybrid approaches, on a more sound theoretical footing. In addition, insights into the relative values of the exchange and correlation weights may be obtained and give a rationale for the values typically obtained by empirical optimization~\cite{2blehybrids_Fromager2011JCP}. One key advantage of considering a linear AC, in place of a generalized path, is that the complementary density-functionals required may be readily derived by the application of uniform coordinate scaling relations to existing standard functionals~\cite{PerSca,YangSca,PRB_Levy_Perdew_Eclambda,LEVY:1985p1777,Joubert:1998p633}, see Refs.~\cite{2blehybrids_Julien,sharkas:044104} for examples of methodologies utilizing this idea. 

Whilst the (generalized) AC provides a legitimization for combining density-functional and wave function methodologies the efficacy of such approaches can only be measured by careful benchmarking. Given the plethora of possible combinations of density-functional and wave function components, and the associated opportunities for error cancellations, assessment of double-hybrid forms is certainly non-trivial. Since each double-hybrid functional expression may be thought of as a model for the AC in this work we take an alternative approach and investigate the quality of the underlying AC integrands. Recently, it has become possible to calculate the ACs for atomic and molecular systems using accurate \textit{ab initio} methodologies~\cite{Colonna:1999p628,Savin:2001p1758,Wu:2003p557,Teale:2009p2020,Teale:2010,Teale:2010b}. Here we will employ the method outlined in Refs.~\cite{Wu:2003p557,Teale:2009p2020,Teale:2010,Teale:2010b}, which utilizes Lieb's definition of the universal density-functional~\cite{LIEB:1983p1813}, to generate benchmark AC integrands for comparison with those generated by two-parameter double-hybrid approaches. 

In this work we analyze the linear AC integrands relevant to
double-hybrid methodologies. We commence in Section~\ref{sec:theory} by
introducing the theory of the linear AC. In particular, we consider the
novel division of the linear AC into segments and their description
using density-functional perturbation theory, which is directly relevant
to double-hybrid approaches. In Section~\ref{sec:CompAC} we introduce
the approximations necessary to practically compute these AC integrands.
Firstly we give a brief overview of the method to determine these
quantities using \textit{ab initio} techniques, which serve as a benchmark in this work. This is followed by a description of the route used to calculate the AC integrands relevant to the $\lambda_1$-B2-PLYP type functionals introduced in Ref.~\cite{2blehybrids_Fromager2011JCP}. The relationship between these two-parameter double-hybrid ACs and the standard B2-PLYP method is then discussed in this context. In Section~\ref{sec:results} we present the calculated ACs for a number of small molecules and van der Waals dimers using these approaches. The results highlight the challenges faced in constructing double-hybrid functionals and the insights provided by our approach are outlined. In Section~\ref{sec:conclusions} we make some concluding remarks and discuss prospects for future work in light of our findings.   

\section{Theory}\label{sec:theory}

In this section we introduce the linear AC and discuss how it may be partitioned into a number of segments. This partitioning of the adiabatic integrand enables the consideration of approaches in which density-functional and wave-function based approaches for describing the integrand can be mixed. The segmented integrand is then expanded through second order in density-functional perturbation theory, providing a rigorous framework for one- and two-parameter double-hybrids.   

\subsection{The linear adiabatic connection}\label{subsec:theory:exlinac}

Let us consider a physical density $n$, which is associated with the ground state wave function, $\Psi$, for the Schr\"{o}dinger equation 
\begin{equation}\label{fully-intSchrodingereq}
\Big(\hat{T}+\hat{W}_{\rm
ee}+\hat{V}\Big)\vert\Psi\rangle=E\vert\Psi\rangle,
\end{equation}
where $\hat{V}=\int d{\bf r}\; v({\bf r})\,\hat{n}({\bf r})$ is the corresponding local potential operator, $\hat{T}$ is the kinetic energy operator and
$\hat{W}_{\rm ee}$ is the two-electron repulsion operator. It is well known that formally exact expressions for the exchange and correlation energies
associated with this density can be obtained when considering a fixed-density linear AC~\cite{LANGRETH:1975p1425,GUNNARSSON:1976p1781,Gunnarsson:1977,LANGRETH:1977p1780,SAVIN:2003p635} between
the non-interacting Kohn--Sham (KS) system and the fully-interacting physical system described by Eq.~(\ref{fully-intSchrodingereq}). 

Introducing a parameter $\lambda$ to modulate the strength of the electronic interactions we may write the auxiliary partially interacting Schr\"odinger equation
\begin{equation}\label{lambda-intSchrodingerACeq}
\Big(
\hat{T}+\lambda\hat{W}_{\rm
ee}+\hat{V}^{\lambda}
\Big)\vert\Psi^\lambda\rangle=\mathcal{E}^\lambda\vert\Psi^\lambda\rangle,
\end{equation}
where the local potential operator 
$\hat{V}^{\lambda}=\int d{\bf r}\; v^{\lambda}({\bf
r})\,\hat{n}({\bf r})$ 
is constructed such that the density constraint 
\begin{equation}\label{dens_constraint}
n_{\Psi^\lambda}({\bf r})=n({\bf r}),\hspace{0.5cm} 0\leq\lambda\leq1,
\end{equation}
is fulfilled. In the $\lambda=1$ limit, $v^{\lambda}({\bf r})$ and
$\Psi^\lambda$ should therefore reduce to the physical $v({\bf r})$ and $\Psi$,
respectively. On the other hand, for $\lambda=0$, the KS potential and
determinant $\Phi^{\rm KS}$ are
recovered. Using the Hellmann-Feynman theorem 
\begin{eqnarray}\label{hellmann_feynman_auxi_ener_exact}
{\displaystyle
\frac{\mathrm{d}\mathcal{E}^\lambda}{\mathrm{d}\lambda}}&=& 
\langle \Psi^{\lambda}\vert
\hat{W}_{\rm ee}\vert\Psi^{\lambda}\rangle
+{\displaystyle
\int \mathrm{d}{\bf r}\; \frac{\partial v^{\lambda}}{\partial \lambda}({\bf
r})\,{n}({\bf r})
},
\end{eqnarray}
the ground-state energy of the physical system, described by Eq.~(\ref{fully-intSchrodingereq}), may be expressed as an integral over the interaction strength on the interval $[0,1]$ 
\begin{eqnarray}\label{DS1H_reduction_lambdaeqzero}
E&=& 
\displaystyle
\int^1_0
\frac{\mathrm{d}\mathcal{E}^\nu}{\mathrm{d}\nu}\;\mathrm{d}\nu+\mathcal{E}^0\nonumber\\
&=&\langle \Phi^{\rm KS}\vert
\hat{T}+\hat{V}\vert\Phi^{\rm KS}\rangle
+E_{\rm H}[n]+\displaystyle
\int^1_0
\mathcal{W}_{\rm xc}^{\nu}[n]\;\mathrm{d}\nu
,
\end{eqnarray}
where $E_{\rm H}[n]=
{1}/{2}\int\!\int{n(\mathbf{r}_1)n(\mathbf{r}_2)}/{r_{12}}\, \mathrm{d}\mathbf{r}_1 \mathrm{d}\mathbf{r}_2
$ is the Hartree energy. The exchange--correlation integrand
$\mathcal{W}_{\rm xc}^{\nu}[n]$ can be further
decomposed into exchange and correlation contributions as follows
\begin{eqnarray}\label{Hxcexactintegrand}
\mathcal{W}_{\rm xc}^{\nu}[n]&=&
\mathcal{W}_{\rm x}^{\nu}[n]+\mathcal{W}_{\rm
c}^{\nu}[n],
\nonumber\\
\mathcal{W}_{\rm x}^{\nu}[n]&=&
\langle\Phi^{\rm KS}\vert\hat{W}_{\rm ee}\vert \Phi^{\rm
KS}\rangle-E_{\rm H}[n]
=
E^{\mbox{\tiny HF}}_{\rm x}[\Phi^{\rm KS}],
\nonumber\\
\mathcal{W}_{\rm c}^{\nu}[n]&=&\langle \Psi^{\nu}\vert
\hat{W}_{\rm ee}\vert\Psi^{\nu}\rangle-
\langle\Phi^{\rm KS}\vert\hat{W}_{\rm ee}\vert\Phi^{\rm
KS}\rangle
,
\end{eqnarray}
where $E^{\mbox{\tiny HF}}_{\rm x}[\Phi^{\rm KS}]$ denotes the HF
exchange energy expression calculated for the KS determinant. 

The expressions of Eq.~(\ref{Hxcexactintegrand}), whilst implicit functionals of the density, $n$, are expressed as explicit functionals of the partially- and non-interacting wave functions, $\Psi^\nu$ and $\Phi^{\rm KS}$ , respectively. Since in this work we wish to distinguish clearly between density-functional and wave-function based expressions for the exchange and correlation integrands we introduce the following notation for the exchange--correlation integrand in terms of explicit density functionals 
\begin{eqnarray}\label{HxcexactintegrandHxcdecomp}
\mathcal{W}_{\rm xc}^{\nu}[n]&=&
 E_{\rm x}[n]+\Delta^{\nu}_{\rm c}[n],
\end{eqnarray}
where $E_{\rm x}[n]$ is the exact KS exchange density functional with a value equal to $E^{\mbox{\tiny HF}}_{\rm x}[\Phi^{\rm KS}]$ in Eq.~(\ref{Hxcexactintegrand}).
The correlation contribution to the integrand as an explicit density functional may be determined by considering the correlation energy of a
partially-interacting system,
\begin{eqnarray}\label{lambdaintC_integration}
{E}^{\lambda}_{\rm c}[n]&=&\displaystyle
\int_0^\lambda
\mathcal{W}_{\rm c}^{\nu}[n]\;\mathrm{d}\nu,
\end{eqnarray}
which can be expressed in terms of the usual correlation
density-functional ($\lambda=1$ limit) by means of
a uniform coordinate scaling in the
density~\cite{PerSca,YangSca,PRB_Levy_Perdew_Eclambda,LEVY:1985p1777},
\begin{eqnarray}\label{scaled_corr_fun}\begin{array} {l}
E^{\lambda}_{\rm c}[n]=\lambda^2E_{\rm
c}[n_{1/\lambda}],\\
\\
n_{1/\lambda}(\mathbf{r})=(1/\lambda)^3n(\mathbf{r}/\lambda).
\end{array}
\end{eqnarray}
Differentiating this expression according to
Eq.~(\ref{lambdaintC_integration}) gives the
well-known result 
\begin{eqnarray}\label{deltaclambdaexpr_scaled_density}
\Delta^{\nu}_{\rm c}[n]&=& 
\mathcal{W}_{\rm c}^{\nu}[n]
={\displaystyle\frac{\partial {E}^{\nu}_{\rm c}[n]}{\partial
\nu}
}
\nonumber\\
&=&
2\nu E_{\rm
c}[n_{1/\nu}]+\nu^2
\displaystyle
\frac{\partial {E}_{\rm c}[n_{1/\nu}]}{\partial \nu}
\end{eqnarray}
for the correlation integrand as an explicit density functional. Again we note that although $\Delta^{\nu}_{\rm c}[n]$ and $\mathcal{W}_{\rm c}^{\nu}[n]$ are formally equivalent we use different notations to emphasize how these quantities are determined. This notational distinction will be helpful in further discussing the segmentation of the adiabatic integrand and the description of these segments by different theories.   

\subsection{Segmentation of the adiabatic integrand}\label{subsec:theory:exlinac:segment}

Practical routes to determine the AC integrands usually proceed either by determining wave
functions that fulfil the density constraint in
Eq.~(\ref{dens_constraint}) and then evaluating the expressions in Eq.~(\ref{Hxcexactintegrand}) or by evaluating the expressions of Eqs.~(\ref{HxcexactintegrandHxcdecomp}) and~(\ref{deltaclambdaexpr_scaled_density}) using an approximate exchange--correlation functional. The former route can be used to provide \textit{ab initio} estimates of the AC integrand, whilst the latter is appropriate to provide estimates corresponding to purely density-functional approaches. 

Since the two-parameter double hybrids examined in this work involve both wave-function and density-functional contributions, we examine a segmentation of the linear AC to combine both approaches. To achieve this we decompose the exact exchange integrand in two parts,
\begin{eqnarray}\label{segmented_EXXintegrand}
\mathcal{{W}}^{\nu}_{\rm x}[n]&=&
E^{\mbox{\tiny HF}}_{\rm x}[\Phi^{\rm KS}]
\times
\mathcal{I}_{[0,\lambda_2[}(\nu)
+
E_{\rm x}[n]
	     \times \mathcal{I}_{[
	    \lambda_2, 1]}(\nu),
\end{eqnarray}
and segment the correlation integrand into three parts,
\begin{eqnarray}\label{exact_integrand_lambda1-lambda2}
\mathcal{W}^{\nu}_{\rm c}[n]
&=&
\Big(\displaystyle
E_{\rm x}[n]
-E^{\mbox{\tiny HF}}_{\rm x}[\Phi^{\rm KS}]
+\Delta^{\nu}_{\rm c}[n]
-\Delta^{\nu}_{\rm c}[n_{\Psi^{\nu}}]
\nonumber\\
&&+
\langle \Psi^{\nu}\vert
\hat{W}_{\rm ee}\vert\Psi^{\nu}\rangle
-E_{\rm Hx}[n_{\Psi^{\nu}}]
\Big)
\times
\mathcal{I}_{[0,\lambda_1[}(\nu)\nonumber\\
&&+
\Big(E_{\rm x}[n]
-E^{\mbox{\tiny HF}}_{\rm x}[\Phi^{\rm KS}]
+\Delta^{\nu}_{\rm c}[n]
-\Delta^{\lambda_1}_{\rm
c}[n_{\Psi^{\lambda_1}}]
     \nonumber\\
&&+
\langle \Psi^{\lambda_1}\vert
\hat{W}_{\rm ee}\vert\Psi^{\lambda_1}\rangle
-E_{\rm Hx}[n_{\Psi^{\lambda_1}}] 
\Big)
 \times    \mathcal{I}_{[\lambda_1,{\lambda_2}[}(\nu)\nonumber\\
&&+\Delta^{\nu}_{\rm c}[n]
	     \times \mathcal{I}_{[
	    \lambda_2, 1]}(\nu),
\end{eqnarray}
where $0\leq\lambda_1\leq\lambda_2\leq1$ and $E_{\rm Hx}[n]=E_{\rm H}[n]+E_{\rm x}[n]$. The indicator function is defined as follows
\begin{eqnarray}\label{step_function}
\mathcal{I}_{A}(\nu)=\left\{
  \begin{array}{l l}
  1   & \quad {\nu}\in A \\
  0   & \quad {\nu}\notin A\\
	      \end{array} \right.
.
\end{eqnarray}
As long as the AC is exact, meaning that the density constraint is
fulfilled, the expressions for the correlation integrand in
Eqs.~(\ref{Hxcexactintegrand}), (\ref{deltaclambdaexpr_scaled_density})
and (\ref{exact_integrand_lambda1-lambda2}) are equivalent. The
advantage of not simplifying further the latter in the first and
second segments, $[0,\lambda_1[$ and $[\lambda_1,\lambda_2[$
respectively, lies in the fact that approximate descriptions of the AC
involving both wave functions and density-functionals can then be formulated and calculated easily by relaxing the density constraint.
Once this constraint is relaxed the first two pairs of terms in the
parentheses relating to the first and second intervals will not compensate anymore, since $n \neq n_{\Psi^\nu}$. However, as we shall see, 
their contributions may be expected to be small under certain conditions.

The first segment of the correlation AC in the interval $[0,\lambda_1[$ has been constructed from the pure
density-functional exchange--correlation expression in Eq.~(\ref{HxcexactintegrandHxcdecomp}), where (i) The HF exchange
based on the KS determinant has been substituted for the exchange density-functional energy according to Eq.~(\ref{segmented_EXXintegrand}) and (ii) The
correlation density-functional integrand $\Delta^{\nu}_{\rm c}[n]$ is
removed by subtracting $\Delta^{\nu}_{\rm c}[n_{\Psi^{\nu}}]$ based on
the wave-function density and then replaced, by
$
\langle \Psi^{\nu}\vert
\hat{W}_{\rm ee}\vert\Psi^{\nu}\rangle
-E_{\rm Hx}[n_{\Psi^{\nu}}]
$. 
Even when the density constraint is relaxed the latter term is the dominant contribution to the correlation integrand in this segment and is purely wave function-dependent. In this
respect, the first segment is predominantly a wave function one. In the
second segment for the interval $[\lambda_1,\lambda_2[$, the correlation integrand varies as $\Delta^{\nu}_{\rm
c}[n]$. The additional wave function
terms simply ensure a continuous transition from the first to the
second segment, which can be referred to as hybrid wave function/density-functional segment.
The final segment $[\lambda_2,1]$ is described within  
density-functional theory.   
As shown in Appendix~\ref{Asec:IntegSegAC}, this partitioning of the
integrand leads, by integration 
over
$[0,1]$, to the exact energy expression 
initially proposed by Fromager~\cite{2blehybrids_Fromager2011JCP}: 
\begin{eqnarray}\label{gsener_lambda12-wf}\begin{array} {l}
E=  
\langle \Psi^{\lambda_1}\vert
\hat{T}+\lambda_2\hat{W}_{\rm ee}+\hat{V}\vert\Psi^{\lambda_1}\rangle
+\overline{E}^{\lambda_1,\lambda_2}_{\rm Hxc}[n_{\Psi^{\lambda_1}}]
,
\end{array}
\end{eqnarray}
where the complementary density-functional contribution equals
\begin{eqnarray}\label{complambda12Hxcfun}\begin{array} {l}
\overline{E}^{\lambda_1,\lambda_2}_{\rm
Hxc}[n]=\overline{E}^{\lambda_1}_{\rm
Hxc}[n]+
\big(\lambda_1-\lambda_2\big)\Big(E_{\rm Hx}[n]+
\Delta^{\lambda_1}_{\rm c}[n]\Big)
,
\end{array}
\end{eqnarray}
with
\begin{eqnarray}\label{Hxcm1decomp}\begin{array} {l}
\overline{E}^{\lambda_1}_{\rm Hxc}[n]=
(1-\lambda_1)E_{\rm Hx}[n]
+
E_{\rm c}[n]-E^{\lambda_1}_{\rm
c}[n]
,\\
\\
\end{array}
\end{eqnarray}
and the wave function $\Psi^{\lambda_1}$ is obtained self-consistently as follows 
\begin{eqnarray}\label{calc_psilambda1}
\Psi^{\lambda_1}&\;\;\leftarrow\;\;&{\displaystyle
 \underset{\Psi}{\rm min}\left\{ \langle \Psi\vert
\hat{T}+\lambda_1\hat{W}_{\rm ee}+\hat{V}\vert\Psi\rangle
+\overline{E}^{\lambda_1}_{\rm Hxc}[n_{\Psi}]\right\}  
}.
\end{eqnarray}

\subsection{Density-functional perturbation theory}\label{subsec:theory:exlinac:ds2dh_sec}

A perturbation expansion of the exact ground-state energy expressed in
Eq.~(\ref{gsener_lambda12-wf})
can be obtained when solving
Eq.~(\ref{calc_psilambda1}) within MP perturbation
theory~\cite{2blehybrids_Fromager2011JCP,2blehybrids_Julien}. By analogy to
the HF approximation, the
zeroth-order wave function $\Phi^{\lambda_1}$ is obtained when
restricting the minimization in Eq.~(\ref{calc_psilambda1}) to single
determinant wave functions $\Phi$: 
\begin{eqnarray}\label{calc_philambda1}
\Phi^{\lambda_1}&\;\;\leftarrow\;\;&{\displaystyle
 \underset{\Phi}{\rm min}\left\{ 
 \langle \Phi\vert
\hat{T}+\lambda_1\hat{W}_{\rm ee}+\hat{V}\vert\Phi\rangle
+\overline{E}^{\lambda_1}_{\rm Hxc}[n_{\Phi}]
\right\}  
}.
\end{eqnarray}
Let us introduce 
a perturbation strength $\alpha$ and the auxiliary
energy 
\begin{eqnarray}\label{newptalpha}\begin{array}{l}
\displaystyle 
{E}^{\alpha,\lambda_1,\lambda_2}
=
{E}^{\alpha,\lambda_1}
-\subsupi{\overline{E}}{Hxc}{\lambda_1}[n_{\Psi^{\alpha,\lambda_1}}]
\\
\\
\hspace{1.9cm}+\alpha\big(\lambda_2-\lambda_1\big)\dfrac{\langle\Psi^{\alpha,\lambda_1}|\subsupi{\hat{W}}{\rm ee}{}|\Psi^{\alpha,\lambda_1}\rangle}{\langle\Psi^{\alpha,\lambda_1}|\Psi^{\alpha,\lambda_1}\rangle}
+
\subsupi{\overline{E}}{\rm
Hxc}{\lambda_1,\lambda_2}[n_{\Psi^{\alpha,\lambda_1}}]
,
\end{array}
\end{eqnarray}
with 
\begin{eqnarray}\label{Ealphamu_var-pt}\begin{array} {l}
{\displaystyle
\Psi^{\alpha,\lambda_1}\;\;\leftarrow\;\;\underset{\Psi}{\rm min}\Big\{ \langle \Psi\vert
\hat{T}+
\lambda_1\subsupi{\hat{U}}{\mbox{\tiny HF}}{}[\Phi^{\lambda_1}]
+
\alpha\lambda_1\supi{\hat{\mathcal{W}}}{\lambda_1}+\hat{V}
\vert\Psi\rangle
}\\
\\
\hspace{2.87cm}
+\overline{E}^{\lambda_1}_{\rm Hxc}[n_{\Psi}]\Big\}  
=E^{\alpha,\lambda_1},
\end{array}
\end{eqnarray}
where $\subsupi{\hat{U}}{\mbox{\tiny HF}}{}[\Phi^{\lambda_1}]$ is the HF potential
operator calculated with $\Phi^{\lambda_1}$, and the perturbation operator
\begin{eqnarray}\label{lrMPpert}\begin{array} {l}
\lambda_1\supi{\hat{\mathcal{W}}}{\lambda_1}=\lambda_1\Big(\subsupi{\hat{W}}{ee}{}-
\subsupi{\hat{U}}{\mbox{\tiny HF}}{}[\Phi^{\lambda_1}]\Big),
\end{array}
\end{eqnarray}
is the scaled fluctuation potential~\cite{2blehybrids_Julien}. 
It is clear, from Eqs.~(\ref{calc_philambda1}) and
(\ref{Ealphamu_var-pt}), that in the $\alpha=0$ limit,
$\Psi^{\alpha,\lambda_1}$ reduces to $\Phi^{\lambda_1}$, while,
according to Eqs.~(\ref{gsener_lambda12-wf}), (\ref{calc_psilambda1}) and
(\ref{newptalpha}), the auxiliary energy becomes, for $\alpha=1$, the {exact}
ground-state energy since $\Psi^{\alpha,\lambda_1}$ reduces 
to $\Psi^{\lambda_1}$. 
We note that the perturbation theory presented in this work 
differs from the one of Sharkas {\it et al.}~\cite{2blehybrids_Julien} by the auxiliary
energy which, in their approach, reduces to $E^{\alpha,\lambda_1}$. Its
perturbation expansion through second order,
\begin{eqnarray}\label{ptalpha_second}\begin{array}{l}
\displaystyle 
{E}^{\alpha,\lambda_1}
={E}^{(0)\lambda_1}+\alpha{E}^{(1)\lambda_1}+\alpha^2\lambda_1^2\subsupi{E}{\mbox{\tiny
MP}}{(2)\lambda_1}+\mathcal{O}(\alpha^3),
\end{array}
\end{eqnarray}
where 
\begin{eqnarray}\label{HF-srDFT-eq2}
{E}^{(0)\lambda_1}+{E}^{(1)\lambda_1}
&=&
 \langle \Phi^{\lambda_1}\vert
\hat{T}+\lambda_1\hat{W}_{\rm ee}+\hat{V}\vert\Phi^{\lambda_1}\rangle
+\overline{E}^{\lambda_1}_{\rm Hxc}[n_{\Phi^{\lambda_1}}],
\end{eqnarray}
will be used in the following. The perturbation
expansion of the wave function, which is deduced from
Eq.~(\ref{Ealphamu_var-pt}), is
however the same. The derivation presented here complements the work of
Fromager~\cite{2blehybrids_Fromager2011JCP}, where perturbation
theory was formulated in terms of an optimized
effective potential (OEP) instead of a density-functional one.
Finally, note the
normalization factor in front of the two-electron interaction expectation
value in 
Eq.~(\ref{newptalpha}), which must be introduced since the intermediate normalization 
condition 
\begin{eqnarray}\label{intnorma_alpha}
\langle
\Phi^{\lambda_1}\vert\Psi^{\alpha,\lambda_1}\rangle=1,
\hspace{0.5cm} 0\leq
\alpha\leq 1,
\end{eqnarray}
will be used.
It has been shown~\cite{pra_MBPTn-srdft,pra_MBPTn-srdft_janos,mp2srdftdmat}
that, in this case, the wave
function can be expanded through second order as follows
\begin{eqnarray}\label{wf-pertalpha}
|\Psi^{\alpha,\lambda_1}\rangle&=&|\Phi^{\lambda_1}\rangle+\alpha\lambda_1|\Psi_{\mbox{\tiny
MP}}^{\rm
(1)\lambda_1}\rangle+\alpha^2|\Psi^{\rm
(2)\lambda_1}\rangle+\mathcal{O}(\alpha^3),
\end{eqnarray}
where $\Psi_{\mbox{\tiny MP}}^{\rm(1)\lambda_1}$ is the analog of the usual MP1 wave
function correction. According to the Brillouin theorem, the density remains unchanged through first
 order, leading
to the following Taylor expansion, through second order, for the
density: 
\begin{eqnarray}\label{density-pert}
n_{\Psi^{\alpha,\lambda_1}}({\bf r})&=&n_{\Phi^{\lambda_1}}({\bf
r})+\alpha^2\delta n^{(2)\lambda_1}({\bf
r})+\mathcal{O}(\alpha^3),
\end{eqnarray}
so that self-consistency effects in Eq.~(\ref{Ealphamu_var-pt}) do
not contribute to the wave function through first
order~\cite{pra_MBPTn-srdft}. Non-zero
contributions actually appear through second order in the wave
function~\cite{mp2srdftdmat}. From the wave function
perturbation expansion in Eq.~(\ref{wf-pertalpha}) and the intermediate
normalization condition, we obtain 
the orthogonality condition
$\langle\Phi^{\lambda_1}\vert\Psi_{\mbox{\tiny MP}}^{(1)\lambda_1}\rangle=0$ and,
as a result, the following Taylor expansion
\begin{eqnarray}\label{srexpvalue_alpha}\begin{array}{l}
\displaystyle
\frac{\left\langle\Psi^{\alpha,\lambda_1}\right|\subsupi{\hat{W}}{\rm ee}{}\left|\Psi^{\alpha,\lambda_1}\right\rangle}{\langle
\Psi^{\alpha,\lambda_1}\vert\Psi^{\alpha,\lambda_1}\rangle}=
\langle\Phi^{\lambda_1}\vert\subsupi{\hat{W}}{\rm ee}{}\vert\Phi^{\lambda_1}\rangle
\\
\\
+2\alpha\lambda_1\langle\Phi^{\lambda_1}\vert\subsupi{\hat{W}}{\rm
ee}{}\vert\Psi_{\mbox{\tiny
MP}}^{\rm
(1)\lambda_1}\rangle
+\mathcal{O}(\alpha^2),
\end{array}
\end{eqnarray}
where, according to Eq.~(\ref{lrMPpert}), the first-order contribution can be rewritten as
\begin{eqnarray}\label{coupling_simple_mp2_exp}
\langle\Phi^{\lambda_1}\vert\subsupi{\hat{W}}{\rm
ee}{}\vert\Psi_{\mbox{\tiny
MP}}^{\rm
(1)\lambda_1}\rangle
&=&  
\langle\Phi^{\lambda_1}\vert\subsupi{\hat{\mathcal{W}}}{
}{\lambda_1}\vert\Psi_{\mbox{\tiny
MP}}^{\rm
(1)\lambda_1}\rangle\nonumber\\
&=& E^{(2)\lambda_1}_{\mbox{\tiny MP}},
\end{eqnarray}
since $\Psi_{\mbox{\tiny MP}}^{(1)\lambda_1}$ contains double excitations only.  
In addition, according to Eq.~(\ref{density-pert}), the 
complement density-functional Hxc energy difference can be expanded through second
order as
\begin{eqnarray}\label{ptexp-srfuncdiff_alpha}\begin{array}{l}
\Big(\subsupi{\overline{E}}{\rm
Hxc}{\lambda_1,\lambda_2}
-\subsupi{\overline{E}}{Hxc}{\lambda_1}\Big)[n_{\Psi^{\alpha,\lambda_1}}]
=
\Big(\subsupi{\overline{E}}{\rm
Hxc}{\lambda_1,\lambda_2}
-\subsupi{\overline{E}}{Hxc}{\lambda_1}\Big)[n_{\Phi^{\lambda_1}}]
\\
\\
{\displaystyle
+\alpha^2
\int d\mathbf{r}\;\left(
\frac{\delta
\overline{E}^{\lambda_1,\lambda_2}_{\rm Hxc}}{\delta
n(\mathbf{r})}
-\frac{\delta
\overline{E}^{\lambda_1}_{\rm Hxc}}{\delta
n(\mathbf{r})}
\right)[n_{\Phi^{\lambda_1}}]\delta
n^{(2)\lambda_1}(\mathbf{r})
}
\\
\\
+\mathcal{O}(\alpha^3).
\end{array}
\end{eqnarray}
As a result, we finally obtain from Eqs.~(\ref{newptalpha}),
(\ref{ptalpha_second}), (\ref{srexpvalue_alpha}) and
(\ref{ptexp-srfuncdiff_alpha}) the following Taylor expansion for the
auxiliary energy
\begin{eqnarray}\label{newptalpha_second}\begin{array}{l}
\displaystyle 
{E}^{\alpha,\lambda_1,\lambda_2}
={E}^{(0)\lambda_1,\lambda_2}+\alpha{E}^{(1)\lambda_1,\lambda_2}+\alpha^2{E}^{(2)\lambda_1,\lambda_2}+\mathcal{O}(\alpha^3),
\end{array}
\end{eqnarray}
where
\begin{eqnarray}\label{newPTexpener_alpha}\begin{array} {l}
{E}^{(0)\lambda_1,\lambda_2}
={E}^{(0)\lambda_1}+
\Big(\subsupi{\overline{E}}{\rm
Hxc}{\lambda_1,\lambda_2}
-\subsupi{\overline{E}}{Hxc}{\lambda_1}\Big)[n_{\Phi^{\lambda_1}}]
,
\\
\\
{E}^{(1)\lambda_1,\lambda_2}
={E}^{(1)\lambda_1}+
\big(\lambda_2-\lambda_1\big) 
\langle\Phi^{\lambda_1}\vert\subsupi{\hat{W}}{\rm ee}{}\vert\Phi^{\lambda_1}\rangle
,
\\
\\
{E}^{(2)\lambda_1,\lambda_2}
=\lambda_1^2\subsupi{E}{\mbox{\tiny MP}}{(2)\lambda_1}
+2\lambda_1\big(\lambda_2-\lambda_1\big)
E^{(2)\lambda_1}_{\mbox{\tiny MP}}
\\
\\
{\displaystyle
+
\int d\mathbf{r}\;\left(
\frac{\delta
\overline{E}^{\lambda_1,\lambda_2}_{\rm Hxc}}{\delta
n(\mathbf{r})}
-\frac{\delta
\overline{E}^{\lambda_1}_{\rm Hxc}}{\delta
n(\mathbf{r})}
\right)[n_{\Phi^{\lambda_1}}]\delta
n^{(2)\lambda_1}(\mathbf{r})
}
.\\
\end{array}
\end{eqnarray}
The exact perturbation expansion of the energy through second order is
then obtained in the $\alpha=1$ limit, which gives, according to
Eq.~(\ref{HF-srDFT-eq2}),
\begin{eqnarray}\label{ds2dhenerexp}
E
&=&  
\langle \Phi^{\lambda_1}\vert
\hat{T}+\lambda_2\hat{W}_{\rm ee}+\hat{V}\vert\Phi^{\lambda_1}\rangle
+\overline{E}^{\lambda_1,\lambda_2}_{\rm Hxc}[n_{\Phi^{\lambda_1}}]
\nonumber\\
&&
+\Big(\lambda_1^2+
2\lambda_1(\lambda_2-\lambda_1)
\Big)E^{(2)\lambda_1}_{\mbox{\tiny
MP}}\nonumber\\  
&&
{\displaystyle
+\int d\mathbf{r}\;\left(
\frac{\delta
\overline{E}^{\lambda_1,\lambda_2}_{\rm Hxc}}{\delta
n(\mathbf{r})}
-\frac{\delta
\overline{E}^{\lambda_1}_{\rm Hxc}}{\delta
n(\mathbf{r})}
\right)[n_{\Phi^{\lambda_1}}]\delta
n^{(2)\lambda_1}(\mathbf{r})
}
\nonumber\\
&&+\ldots
\end{eqnarray}
This energy expression is associated with the segmented correlation
integrand in Eq.~(\ref{exact_integrand_lambda1-lambda2}) whose perturbation expansion through second order
is obtained when decomposing the auxiliary energy as follows 
\begin{eqnarray}\label{segments_auxener_pt}
{E}^{\alpha,\lambda_1,\lambda_2}
&=&
{E}^{\alpha,0,0}+
\int^{\lambda_1}_0
\frac{\mathrm{d}{E}^{\alpha,\nu,\nu}}{\mathrm{d}\nu} \mathrm{d}\nu+
\int_{\lambda_1}^{\lambda_2}
\frac{\mathrm{d}{E}^{\alpha,\lambda_1,\nu}}{\mathrm{d}\nu} \mathrm{d}\nu
\nonumber\\
&=&
\langle \Phi^{\rm KS}\vert
\hat{T}+\hat{V}\vert\Phi^{\rm KS}\rangle
+E_{\rm H}[n]+\displaystyle
\int^1_0
\mathcal{{W}}^{\nu}_{\rm
x}[n]\;\mathrm{d}\nu
\nonumber\\
&&+\int^1_0
\mathcal{{W}}^{\alpha,\lambda_1,\lambda_2,\nu}_{\rm c}\;\mathrm{d}\nu,
\end{eqnarray}
where the exchange integrand is expressed as in
Eq.~(\ref{segmented_EXXintegrand}) and, in the $\alpha=1$ limit,
$\mathcal{{W}}^{\alpha,\lambda_1,\lambda_2,\nu}_{\rm c}$ reduces to the
exact correlation integrand $\mathcal{W}^{\nu}_{\rm c}[n]$. The particular case $\lambda_1=\lambda_2=0$ corresponds to the exact
KS theory:
\begin{eqnarray}\label{Eauxalpha00}
{E}^{\alpha,0,0}
&=&
\langle \Phi^{\rm KS}\vert
\hat{T}+\hat{V}\vert\Phi^{\rm KS}\rangle
+E_{\rm H}[n]+\displaystyle
\int^1_0
\Big(E_{\rm x}[n]+\Delta^{\nu}_{\rm c}[n]\Big)\;\mathrm{d}\nu
.
\end{eqnarray}
As shown in Appendix~\ref{Asec:PTexpSegAC}, the additional terms in the
right-hand side of Eq.~(\ref{segments_auxener_pt}) (first line) introduce many-body perturbation theory
corrections to the correlation integrand, in the first and second
segments, which leads to the following perturbation
expansion when considering the $\alpha=1$ limit: 
\begin{eqnarray}\label{ds2dh_integrand}
\mathcal{W}^{\nu}_{\rm c}[n]&=&
\Big(\displaystyle
E_{\rm x}[n]
-E^{\mbox{\tiny HF}}_{\rm x}[\Phi^{\rm KS}]
+\Delta^{\nu}_{\rm c}[n]
-\Delta^{\nu}_{\rm
c}[n_{\Phi^{\nu}}]\nonumber\\
&&+\langle \Phi^{\nu}\vert
\hat{W}_{\rm ee}\vert\Phi^{\nu}\rangle
+2\nu E^{(2)\lambda_1}_{\mbox{\tiny MP}}
-E_{\rm Hx}[n_{\Phi^{\nu}}]
\Big)
\times
\mathcal{I}_{[0,\lambda_1[}(\nu)\nonumber\\
&&+\Bigg(E_{\rm x}[n]
-E^{\mbox{\tiny HF}}_{\rm x}[\Phi^{\rm KS}]
+\Delta^{\nu}_{\rm c}[n]
\nonumber
\\
&&
-\Delta^{\lambda_1}_{\rm
c}[n_{\Phi^{\lambda_1}}]
-\int d\mathbf{r}\;
{\delta{\Delta}^{\lambda_1}_{\rm c}}/{\delta
n(\mathbf{r})}[n_{\Phi^{\lambda_1}}]
\delta
n^{(2)\lambda_1}(\mathbf{r})
\nonumber\\
&&+\langle \Phi^{\lambda_1}\vert
\hat{W}_{\rm ee}\vert\Phi^{\lambda_1}\rangle
+
2\lambda_1 E^{(2)\lambda_1}_{\mbox{\tiny MP}}
\nonumber
\\
&&
-E_{\rm Hx}[n_{\Phi^{\lambda_1}}]
-\int d\mathbf{r}\;
{\delta{E}_{\rm Hx}}/{\delta
n(\mathbf{r})}[n_{\Phi^{\lambda_1}}]
\delta
n^{(2)\lambda_1}(\mathbf{r})
\Bigg)
\nonumber
\\
&&+\times\mathcal{I}_{[\lambda_1,{\lambda_2}[}(\nu)
\nonumber\\
&&+\Delta^{\nu}_{\rm c}[n]
	   \times \mathcal{I}_{
	    [\lambda_2, 1]}(\nu)
	    \nonumber\\
	    &&+\ldots
\end{eqnarray}
Note that, in the first segment, the MP2 contribution has been
linearized for convenience. Nevertheless, after integration, the exact
MP2 correlation energy is recovered (see Eq.~(\ref{integrationmp2term})). From the exact integrand expression in
Eq.~(\ref{exact_integrand_lambda1-lambda2}) it is clear that, in both first and second segments, the wave function has been expanded
in MP perturbation theory through second order. The third segment, which is the pure DFT part, is not
modified by the perturbation theory treatment. 

Interestingly, the
second-order correction to the density only appears in the second
segment. This is due to the fact that, in the particular case
$\lambda_1=\lambda_2$ (that is when the second segment disappears), the
$2n+1$ rule is
fulfilled~\cite{pra_MBPTn-srdft_janos,janos_2nplus1rule_MBPTn-srdft}. As a result, second-order corrections to the
density (and therefore to the wave function) are absent from the
second-order correction to the energy. In the general case, where
$\lambda_1\neq\lambda_2$, the second-order corrections to the density
introduce a discontinuity at $\lambda_1$ in the correlation integrand.
This would in principle disappear when expanding the integrand in the two first
segments to infinite order, provided that the perturbation theory converges
smoothly of course, which might not be the case in
practice~\cite{mp2srdftdmat}. Finally we stress that, in the first
two segments, all quantities calculated with $\Phi^{\nu}$ $(\nu>0)$
correspond to correlation effects as defined in KS-DFT. In other words
orbital relaxations which make $\Phi^{\nu}$ differ from $\Phi^{\rm KS}$
contribute to the correlation energy, exactly like in second-order
G\"{o}rling-Levy perturbation theory (GL2)~\cite{glpt2}. 

\section{Computing the AC}\label{sec:CompAC}

In this section we introduce the methods used to calculate the AC integrands. The most accurate approach calculates the exchange and correlation integrands introduced in Sec.~\ref{subsec:theory:exlinac} by means of {\it ab initio} methods.  The integrands calculated in this manner will be used to serve as a benchmark for more approximate approaches. To analyze practical double-hybrid methods based on the $\lambda_1$-B2-PLYP  methods, introduced in Ref.~\cite{2blehybrids_Fromager2011JCP}, an approximate formulation of the second-order density-functional perturbation theory presented in Sec.~\ref{subsec:theory:exlinac:ds2dh_sec} is considered. For comparison a similar approach is also applied to determine an AC for the conventional B2-PLYP functional.


\subsection{ab initio estimates of the AC integrand}\label{subsec:theory:approxac:abinit}

To calculate accurate \textit{ab initio} estimates of the AC we employ the methodology in Refs.~\cite{Wu:2003p557,Teale:2009p2020,Teale:2010,Teale:2010b}. Following Lieb~\cite{LIEB:1983p1813}, we write the auxiliary energy as
\begin{equation}
\mathcal{E}^\lambda[v] = \inf_n \left[ F^\lambda[n] + \int v(\mathbf{r}) n(\mathbf{r}) \mathrm{d}\mathbf{r} \right], 
\label{ELieb}
\end{equation}
which gives the ground-state energy for the auxiliary Hamiltonian
\begin{equation}
\hat{H}^{\lambda}[v]=\hat{T}+\lambda\hat{W}_{\rm ee}+\int \mathrm{d}{\bf r}\; v({\bf r})\,\hat{n}({\bf r}). 
\end{equation}
The universal density functional $F^\lambda[n]$ can be expressed as a Legendre--Fenchel transform (convex conjugate) to the ground-state auxiliary energy $\mathcal{E}^\lambda[v]$,
\begin{equation}
F^\lambda[n] = \sup_v \left[ \mathcal{E}^\lambda[v] - \int v(\mathbf{r})
n(\mathbf{r}) \mathrm{d}\mathbf{r} \right], 
\label{FLieb}
\end{equation}
where the maximization is over a complete vector space of potentials. See Refs.~\cite{Eschrig,Kutzelnigg:2006,vanLeeuwen:2003} for reviews of this approach to DFT. In the present work we employ \textit{ab initio} approaches to calculate $\mathcal{E}^\lambda[v]$ accurately and hence determine the functional $F^\lambda[n]$ accurately. We note that even for approximate theories in finite basis sets where $\mathcal{E}^\lambda[v]$ may not be guaranteed to be concave in $v$ the functional of Eq.~(\ref{FLieb}) may still be constructed in a well defined manner, being conjugate to the concave envelope of $\mathcal{E}^\lambda[v]$ at a given level of theory, which is denoted $\bar{\mathcal{E}}^\lambda[v]$. The concave envelope provides an upper bound, $\bar{\mathcal{E}}^\lambda[v] \geq \mathcal{E}^\lambda[v]$, with equality when $\mathcal{E}^\lambda[v]$ is concave in $v$. In the limit of a full configuration-interaction treatment of correlation and a complete one-electron basis set the exact universal-density functional is recovered. 

For practical calculations we employ the algorithm proposed in Ref.~\cite{Wu:2003p557} and implemented in Refs.~\cite{Teale:2009p2020,Teale:2010,Teale:2010b} for arbitrary interaction strengths. The key aspect of this approach is to introduce an expansion of the potential
\begin{equation}
v_{\mathbf{b}}(\mathbf{r}) = v_{\text{ext}}(\mathbf{r}) + (1-\lambda)v_{\text{ref}}(\mathbf{r}) + \sum_t b_t g_t(\mathbf{r}),
\end{equation}  
which allows for the use of analytic derivatives in quasi-Newton approaches to perform the optimization of Eq.~(\ref{FLieb}) and determine the potential expansion coefficients $\{ b_t \}$. This opens up the possibility to perform calculations on molecular systems to complement earlier approaches applicable to atomic species~\cite{Colonna:1999p628,Savin:2001p1758}. Here we use the second order optimization scheme detailed in Refs.~\cite{Wu:2003p557} with a truncated singular value cutoff of $10^{-6}$ on the Hessian and a gradient norm tolerance of $10^{-6}$. The Fermi--Amaldi potential is employed for $v_{\text{ref}}$ and we use the same basis set $\{ g_t \}$ for the potential expansion as is used for the molecular orbitals. To determine the densities $n$ we use the Lagrangian method of Helgaker and J{\o}rgensen~\cite{helg:1989:lag,jorg:1988:lag,koch:1990:lag,hald:2003:lag}, where required, to calculate the relaxed density matrices. For the H$_2$ molecule we perform calculations at the FCI level to determine $\mathcal{E}^\lambda[v]$, $n$, and $F^\lambda[n]$. For all other species considered in this work we employ the coupled-cluster singles-doubles and perturbative triples [CCSD(T)]~\cite{CCSDT} method with all electrons correlated. All calculations are performed with a modified version of the DALTON2011 program~\cite{DALTON}.

To make the link to the AC we note that the Lieb functional of Eq~(\ref{FLieb}) is equivalent to the Levy--Lieb constrained search functional for canonical ensembles~\cite{LIEB:1983p1813}
\begin{equation}
F^\lambda[n] = \min_{\hat{\gamma}\rightarrow n} \text{Tr}
\hat{H}^\lambda[0]\hat{\gamma} = \text{Tr}
\hat{H}^\lambda[0]\hat{\gamma}^{\lambda,n}, 
\label{CSLieb}
\end{equation}
where the minimization is over all density matrices with density $n$ and ${\gamma}^{\lambda,n}$ is the minimizing density matrix. The interacting functional $F^\lambda[n]$ can be related to the non-interacting functional via
\begin{equation}
F^\lambda[n] = F^0[n] + \int_0^\lambda
\frac{\mathrm{d}F^{\nu}[n]}{\mathrm{d}\nu} \mathrm{d} \nu.
\end{equation}
Identifying $F^0[n]$ with $T_\text{s}[n]$, evaluating $\frac{\mathrm{d}}{\mathrm{d}\nu}F^{\nu}[n]$ by differentiating Eq.~(\ref{CSLieb}) and employing the Hellmann-Feynman theorem we obtain the usual AC expression
\begin{equation}
F^\lambda[n] = T_\text{s}[n] + \int_0^\lambda
\mathcal{W}_{\text{Hxc}}^{\nu}[n] \mathrm{d} \nu,
\end{equation}
where the AC integrand can be decomposed into
\begin{equation}
 \int_0^\lambda \mathcal{W}_{\text{Hxc}}^{\nu}[n] \mathrm{d} \nu =
 \lambda E_{\text{H}}[n] + \lambda E_\text{x}[n] +
 E_{\text{c}}^{\lambda}[n],
\end{equation}  
where $E_{\text{H}}[n]$ is the classical Coulomb energy.
The exchange energy component is given by
\begin{equation}\label{abinitio_EXX}
E_\text{x}[n]=\text{Tr} \hat{W}_{\rm ee}\hat{\gamma}^{0,n} -
E_{\text{H}}[n],
\end{equation}
and the correlation energy of Eq.~(\ref{lambdaintC_integration}) can be calculated using the correlation integrand
\begin{equation}
\mathcal{W}_{\text{c}}^{\nu}[n] = \text{Tr} \hat{W}_{\rm ee}(\hat{\gamma}^{\nu,n} - \hat{\gamma}^{0,n}).
\end{equation}
Each of the energy components and their corresponding AC integrands can be calculated at each interaction strength following the optimization of Eq.~(\ref{FLieb}), thereby mapping out the AC. 

\subsection{$\lambda_1$-B2-PLYP double hybrid integrand}\label{subsec:theory:approxac:lam1B2PLYP}

We consider in this section an approximate formulation of the
density-functional perturbation theory derived in
Sec.~\ref{subsec:theory:exlinac:ds2dh_sec} where (i) the
energy expansion is truncated at second order (ii) Becke exchange and LYP
correlation density-functionals are used (iii) density scaling in 
the LYP correlation functional is neglected (iv) second-order
corrections to the density are neglected. The
$\lambda_1$-B2-PLYP energy~\cite{2blehybrids_Fromager2011JCP}
is thus recovered from Eq.~(\ref{ds2dhenerexp}),
\begin{eqnarray}\label{connection_ds2dh_conv2dh}
\tilde{E}^{\lambda_1,\lambda_2}
&=&
\langle \tilde{\Phi}^{\lambda_1}\vert
\hat{T}+\hat{V}\vert\tilde{\Phi}^{\lambda_1}\rangle
+
{E}_{\rm
H}[n_{\tilde{\Phi}^{\lambda_1}}]
+\lambda_2E^{\mbox{\tiny HF}}_{\rm x}[\tilde{\Phi}^{\lambda_1}]
\nonumber\\
&&
+\big(1-\lambda_2\big)
{E}^{\mbox{\tiny B}}_{\rm
x}[n_{\tilde{\Phi}^{\lambda_1}}]
+\Big(1-\lambda_1\big(2\lambda_2-\lambda_1\big)\Big)
{E}^{\mbox{\tiny
LYP}}_{\rm
c}
[n_{\tilde{\Phi}^{\lambda_1}}]
\nonumber\\
&&+\lambda_1\big(2\lambda_2-\lambda_1\big)
\tilde{E}^{(2)\lambda_1}_{\mbox{\tiny
MP}}  
,
\end{eqnarray}
where, according to Eq.~(\ref{calc_philambda1}), the orbitals are computed as follows:
\begin{eqnarray}\label{lambda1_orb_calc}
\tilde{\Phi}^{\lambda_1}&\;\;\leftarrow\;\;& \underset{\Phi}{\rm min} \Bigg\{ \langle \Phi\vert
\hat{T}+\hat{V}
\vert\Phi\rangle
+ 
{E}_{\rm
H}[n_{\Phi}]
+
\lambda_1 
E^{\mbox{\tiny HF}}_{\rm x}[\Phi]
\nonumber\\
&&
\hspace{1cm}+\big(1-\lambda_1\big){E}^{\mbox{\tiny B}}_{\rm
x}[n_{\Phi}]
+\big(1-\lambda_1^2\big){E}^{\mbox{\tiny LYP}}_{\rm
c}[n_{\Phi}]
\Bigg\}.
\end{eqnarray}
The $\lambda_1$-B2-PLYP energy expression is formally identical to the conventional
B2-PLYP one. The fractions of HF exchange $a_{\rm x}$ and MP2 correlation
energy $a_{\rm c}$ can be identified as
\begin{eqnarray}\label{axac}\begin{array} {l}
a_{\rm x}=\lambda_2
\\
\\
a_{\rm c}=\lambda_1(2\lambda_2-\lambda_1)
\\
\end{array}
\;\;\;\longleftrightarrow\;\;\;
\begin{array} {l}
\lambda_1=a_{\rm x}-\sqrt{a^2_{\rm x}-a_{\rm c}}
\\
\\
\lambda_2=a_{\rm x}
\end{array},
\end{eqnarray}
as long as the condition $a_{\rm c}\leq a^2_{\rm x}$ is fulfilled, which
is usually the case in conventional one- and two-parameter double
hybrids~\cite{2blehybrids_Fromager2011JCP}. In the spirit of
Eq.~(\ref{segments_auxener_pt}), the $\lambda_1$-B2-PLYP energy can be
rewritten in terms of an exchange--correlation
integrand,
\begin{eqnarray}\label{segments_lambda1b2plyp}
\tilde{E}^{\lambda_1,\lambda_2}
&=&\tilde{E}^{0,0}+
\int^{\lambda_1}_0
\frac{\mathrm{d}\tilde{E}^{\nu,\nu}}{\mathrm{d}\nu} \mathrm{d}\nu+
\int_{\lambda_1}^{\lambda_2}
\frac{\mathrm{d}\tilde{E}^{\lambda_1,\nu}}{\mathrm{d}\nu}
\mathrm{d}\nu
\nonumber\\
&=& 
\langle\tilde{\Phi}^0 \vert
\hat{T}+\hat{V}\vert\tilde{\Phi}^0\rangle
+{E}_{\rm H}[n_{\tilde{\Phi}^0}]
+\displaystyle
\int^1_0
\mathcal{\tilde{W}}^{\lambda_1,\lambda_2,\nu}_{\rm xc}
\mathrm{d}\nu
,
\end{eqnarray}
where the KS-BLYP density $n_{\tilde{\Phi}^0}$ used as reference is
recovered 
in the particular case $\lambda_1=\lambda_2=0$. 
The corresponding
KS-BLYP energy can be expressed as 
\begin{eqnarray}\label{Eblyp}
\tilde{E}^{0,0}&=& 
\langle\tilde{\Phi}^0 \vert
\hat{T}+\hat{V}\vert\tilde{\Phi}^0\rangle
+{E}_{\rm
H}[n_{\tilde{\Phi}^0}]+
\displaystyle
\int^1_0
\bigg({E}^{\mbox{\tiny B}}_{\rm
x}[n_{\tilde{\Phi}^0}]
+2\nu
{E}^{\mbox{\tiny
LYP}}_{\rm
c}[n_{\tilde{\Phi}^0}]
\bigg)
\mathrm{d}\nu
,
\end{eqnarray}
where the uniform coordinate scaling in
the LYP correlation integrand 
\begin{eqnarray}\label{deltaclambdaexpr_scaled_density_blyp}
\Delta^{\mbox{\tiny LYP},\nu}_{\rm c}[n]&=& 
2\nu E^{\mbox{\tiny LYP}}_{\rm
c}[n_{1/\nu}]+\nu^2
\displaystyle
\frac{\partial {E}^{\mbox{\tiny LYP}}_{\rm c}[n_{1/\nu}]}{\partial \nu}
,
\end{eqnarray}
has been neglected. 
By analogy with
Eq.~(\ref{segmented_EXXintegrand}),
we define the $\lambda_1$-B2-PLYP exchange integrand  
in terms of the HF and Becke exchange energies, both computed with the
KS-BLYP determinant $\tilde{\Phi}^{0}$, as follows   
\begin{eqnarray}\label{lambda1b2lypXCintegranddef2}
\mathcal{\tilde{W}}^{\lambda_2,\nu}_{\rm
x}
&=&
E^{\mbox{\tiny HF}}_{\rm x}[\tilde{\Phi}^{0}]
\times
\mathcal{I}_{[0,\lambda_2[}(\nu)
+E^{\mbox{\tiny B}}_{\rm x}[n_{\tilde{\Phi}^0}]
	     \times \mathcal{I}_{[
	    \lambda_2, 1]}(\nu).
\end{eqnarray}
The associated correlation integrand 
\begin{eqnarray}\label{lambda1b2lypXCintegranddef}
\mathcal{\tilde{W}}^{\lambda_1,\lambda_2,\nu}_{\rm c}
&=& 
\mathcal{\tilde{W}}^{\lambda_1,\lambda_2,\nu}_{\rm xc}
-
\mathcal{\tilde{W}}^{\lambda_2,\nu}_{\rm
x},
\end{eqnarray}
can then be deduced from Eqs.~(\ref{segments_lambda1b2plyp}) and
(\ref{Eblyp}), like in 
Sec.~\ref{subsec:theory:exlinac:ds2dh_sec}. 
Since 
\begin{eqnarray}\label{segments_lambda1b2plyp_3rdpart}
\frac{\mathrm{d}\tilde{E}^{\lambda_1,\nu}}{\mathrm{d}\nu} &=&
E^{\mbox{\tiny HF}}_{\rm x}[\tilde{\Phi}^{\lambda_1}]
-{E}^{\mbox{\tiny B}}_{\rm x}[n_{\tilde{\Phi}^{\lambda_1}}]
-2\lambda_1{E}^{\mbox{\tiny LYP}}_{\rm
c}[n_{\tilde{\Phi}^{\lambda_1}}]
\nonumber\\
&&
+2\lambda_1\tilde{E}^{(2)\lambda_1}_{\mbox{\tiny
MP}},
\end{eqnarray}
we finally obtain the following expression for the $\lambda_1$-B2-PLYP correlation integrand 
\begin{eqnarray}\label{lambda1b2plyp_Cintegrand}
\mathcal{\tilde{W}}^{\lambda_1,\lambda_2,\nu}_{\rm c}
&=&
\Big(E^{\mbox{\tiny B}}_{\rm x}[n_{\tilde{\Phi}^0}]
-E^{\mbox{\tiny B}}_{\rm x}[n_{\tilde{\Phi}^{\nu}}]
+E^{\mbox{\tiny HF}}_{\rm x}[\tilde{\Phi}^{\nu}]
-E^{\mbox{\tiny HF}}_{\rm x}[\tilde{\Phi}^{0}]\nonumber\\
&&+2\nu\tilde{E}^{(2)\lambda_1}_{\mbox{\tiny MP}}
+2\nu \big(E^{\mbox{\tiny
LYP}}_{\rm c}[n_{\tilde{\Phi}^0}]
-E^{\mbox{\tiny
LYP}}_{\rm
c}[n_{\tilde{\Phi}^{\nu}}]\big) 
\Big)
\times
\mathcal{I}_{[0,\lambda_1[}(\nu)\nonumber\\
&&+
\Big(
\mathcal{\tilde{W}}^{\lambda_1,\lambda_2,\lambda^-_1}_{\rm c}
+2(\nu-\lambda_1) E^{\mbox{\tiny LYP}}_{\rm c}[n_{\tilde{\Phi}^0}]
\Big)
 \times    \mathcal{I}_{[\lambda_1,{\lambda_2}[}(\nu)\nonumber\\
&&+
2\nu E^{\mbox{\tiny LYP}}_{\rm c}[n_{\tilde{\Phi}^0}]
	     \times \mathcal{I}_{[
	    \lambda_2, 1]}(\nu),
\end{eqnarray}
where $\lambda^-_1$ means $\lambda\underset{\lambda <
\lambda_1}{\longrightarrow}\lambda_1$.
From Eq.~(\ref{lambda1b2plyp_Cintegrand}) we note that the $\lambda_1$-B2-PLYP correlation integrand is
continuous in $\lambda_1$, even though approximate wave function and
density-functionals are used.  
Qualitatively the behaviour of the AC can be understood by neglecting the variation of all terms depending implicitly on $\nu$:
\begin{eqnarray}\label{lambda1b2plyp_qualitativeCintegrand}
\mathcal{\tilde{W}}^{\lambda_1,\lambda_2,\nu}_{\rm c}
&\sim&
2\nu\tilde{E}^{(2)\lambda_1}_{\mbox{\tiny MP}}
\times
\mathcal{I}_{[0,\lambda_1[}(\nu)\nonumber\\
&&+
2\nu E^{\mbox{\tiny LYP}}_{\rm c}[n_{\tilde{\Phi}^0}]
 \times    \mathcal{I}_{[\lambda_1,{\lambda_2}[}(\nu)\nonumber\\
&&+
2\nu E^{\mbox{\tiny LYP}}_{\rm c}[n_{\tilde{\Phi}^0}]
	     \times \mathcal{I}_{[
	    \lambda_2, 1]}(\nu).
\end{eqnarray}
In the first segment $[0,\lambda_1[$, the slope of the
$\lambda_1$-B2-PLYP AC curve is dominated
by the MP2 correlation energy of the auxiliary $\lambda_1$-interacting
system. On the other hand, in the other two segments, the conventional
LYP correlation energy dominates the slope. 
Curvature could be
introduced into the approximate ACs by considering higher-order MP terms and introducing
density scaling effects. In this work we do not consider higher order perturbation theory energies, however,
the effects of density scaling will be investigated further in Sec.~\ref{sec:results}. For that
purpose we define from Eq.~(\ref{ds2dh_integrand})
a $\lambda_1$ density-scaled B2-PLYP ($\lambda_1$-DS-B2-PLYP)
correlation integrand:
\begin{eqnarray}\label{DSlambda1b2plyp_Cintegrand}
\mathcal{\tilde{W}}^{\lambda_1,\lambda_2,\nu}_{\rm c,\mbox{\tiny DS}}&=&
\Big(
\mathcal{\tilde{W}}^{\lambda_1,\lambda_2,\nu}_{\rm c}
+\big(
\Delta^{\mbox{\tiny LYP},\nu}_{\rm c}[n_{\tilde{\Phi}^0}]
-
2\nu E^{\mbox{\tiny LYP}}_{\rm c}[n_{\tilde{\Phi}^0}]\big)\nonumber\\
&&-\big(\Delta^{\mbox{\tiny LYP},\nu}_{\rm c}[n_{\tilde{\Phi}^\nu}]
-2\nu 
E^{\mbox{\tiny
LYP}}_{\rm
c}[n_{\tilde{\Phi}^{\nu}}]\big) 
\Big)
\times
\mathcal{I}_{[0,\lambda_1[}(\nu)\nonumber\\
&&+
\Big(
\mathcal{\tilde{W}}^{\lambda_1,\lambda_2,\lambda^-_1}_{\rm c,\mbox{\tiny
DS}}
+\Delta^{\mbox{\tiny LYP},\nu}_{\rm c}[n_{\tilde{\Phi}^0}]
-\Delta^{\mbox{\tiny LYP},\lambda_1}_{\rm c}[n_{\tilde{\Phi}^0}]
\Big)
 \times    \mathcal{I}_{[\lambda_1,{\lambda_2}[}(\nu)\nonumber\\
&&+
\Delta^{\mbox{\tiny LYP},\nu}_{\rm c}[n_{\tilde{\Phi}^0}]
	     \times \mathcal{I}_{[
	    \lambda_2, 1]}(\nu),
\end{eqnarray}
where the density-scaled LYP correlation integrand is defined in
Eq.~(\ref{deltaclambdaexpr_scaled_density_blyp}). 
Note that the second-order corrections to the density have been neglected.
As a result, the approximate $\lambda_1$-DS-B2-PLYP correlation integrand
remains continuous in $\lambda_1$. Moreover, for simplicity, the
orbitals used in $\lambda_1$-DS-B2-PLYP and $\lambda_1$-B2-PLYP schemes
are the same, which means that density scaling has not been taken into
account in the self-consistent calculation of the orbitals.

The integrands of Eqs.~(\ref{lambda1b2plyp_Cintegrand}) and
(\ref{DSlambda1b2plyp_Cintegrand}) provide an approximate
description of the AC for which the reference density is the KS-BLYP one.
According to Eq.~(\ref{lambda1_orb_calc}) and
Ref.~\cite{2blehybrids_Fromager2011JCP}, the local potential $v^{\nu}$, which ensures that the density
constraint on the $\nu$-interacting wave function is fulfilled, is approximated here by
\begin{eqnarray}\label{lam1hybDFTeqpot}
v^\nu(\mathbf{r})&\rightarrow& v(\mathbf{r})+
{\displaystyle
(1-\nu)\Bigg(\frac{\delta{{E}_{\rm H}}[n_{\tilde{\Phi}^{\nu}}]}{\delta
n(\mathbf{r})}
}
+
\frac{\delta{{E}^{\mbox{\tiny B}}_{\rm x}}[n_{\tilde{\Phi}^{\nu}}]}{\delta
n(\mathbf{r})}
\Bigg)
+(1-\nu^2)\frac{\delta{{E}^{\mbox{\tiny LYP}}_{\rm c}}[n_{\tilde{\Phi}^{\nu}}]}{\delta
n(\mathbf{r})}.
\end{eqnarray}
Due to the Brillouin theorem the zeroth-order density $n_{\tilde{\Phi}^{\nu}}$
remains unchanged through first order and, as illustrated by the
$\lambda_1$-B2-PLYP AC curves in Sec.~\ref{sec:results}, it does not
vary significantly on the segment $[0,\lambda_1[$. 

\subsection{Conventional B2-PLYP double hybrid AC integrand}\label{subsec:theory:approxac:convB2PLYP}

In conventional B2-PLYP calculations, the energy is calculated as follows 
\begin{eqnarray}\label{Conv2H}
\overline{E}^{a_{\rm x},a_{\rm c}}&=&
\langle \overline{\Phi}^{a_{\rm x},a_{\rm c}}\vert
\hat{T}+\hat{V}\vert\overline{\Phi}^{a_{\rm
x},a_{\rm c}}\rangle
+
{E}_{\rm
H}[n_{\overline{\Phi}^{a_{\rm x},a_{\rm c}}}]
+
a_{\rm x}
E^{\mbox{\tiny HF}}_{\rm x}[\overline{\Phi}^{a_{\rm x},a_{\rm c}}]
\nonumber\\
&&
+\big(1-a_{\rm x}\big){E}^{\mbox{\tiny B}}_{\rm
x}[n_{\overline{\Phi}^{a_{\rm x},a_{\rm c}}}]
+\big(1-a_{\rm c}\big){E}^{\mbox{\tiny LYP}}_{\rm
c}[n_{\overline{\Phi}^{a_{\rm x},a_{\rm c}}}]
\nonumber\\
&&
+
a_{\rm c}
\overline{E}^{(2)a_{\rm x},a_{\rm c}}_{\mbox{\tiny
MP}}.
\end{eqnarray}
This expression is formally identical to the
$\lambda_1$-B2-PLYP energy, provided that the condition $a_{\rm c}\leq a^2_{\rm
x}$ is fulfilled, and so the mapping in Eq.~(\ref{axac}) between ($a_{\rm x}$,
$a_{\rm c}$) and ($\lambda_1,\lambda_2$) exists. The difference between the numerical values of the B2-PLYP and $\lambda_1$-B2-PLYP energies lies in
the computation of the orbitals used, where the two parameters $a_{\rm x}$
and $a_{\rm c}$, instead of one, like in the $\lambda_1$-B2-PLYP scheme
(see Eq.~(\ref{lambda1_orb_calc})),
are involved:
\begin{eqnarray}\label{Convb2plyp_calc_orb}
\overline{\Phi}^{a_{\rm x},a_{\rm c}}&\;\;\leftarrow\;\;& \underset{\Phi}{\rm min} \Bigg\{ \langle \Phi\vert
\hat{T}+\hat{V}
\vert\Phi\rangle
+ 
{E}_{\rm
H}[n_{\Phi}]
+
a_{\rm x}
E^{\mbox{\tiny HF}}_{\rm x}[\Phi]
\nonumber\\
&&
\hspace{1cm}+\big(1-a_{\rm x}\big){E}^{\mbox{\tiny B}}_{\rm
x}[n_{\Phi}]
+\big(1-a_{\rm c}\big){E}^{\mbox{\tiny LYP}}_{\rm
c}[n_{\Phi}]
\Bigg\}.
\end{eqnarray}
As a consequence, there is an ambiguity in the way the correlation
integrand should be defined for B2-PLYP. Indeed, the B2-PLYP energy
expression cannot be derived rigorously, from the density-functional
perturbation theory in Sec.~\ref{subsec:theory:exlinac:ds2dh_sec}, as long
as the orbitals are calculated according to
Eq.~(\ref{Convb2plyp_calc_orb}). In this case, the
Brillouin theorem cannot be applied~\cite{2blehybrids_Fromager2011JCP}, which is an
important difference with the 
$\lambda_1$-B2-PLYP scheme, so that single
excitation contributions to the double hybrid energy should in
principle be considered. Nevertheless, it is interesting for analysis
purposes to construct analytically a B2-PLYP integrand that can be compared to the
$\lambda_1$-B2-PLYP one. The simple
segmentation
\begin{eqnarray}\label{convb2plyp_othersegmentation}
\overline{E}^{a_{\rm x},a_{\rm c}}&=&
\overline{E}^{0,0}
+
\displaystyle
\int^{a_{\rm x}}_0
\;
\frac{\mathrm{d}}{\mathrm{d}\nu}\overline{E}^{\nu,\nu^2}
\mathrm{d}\nu
\displaystyle
-\int^{a^2_{\rm x}}_{a_{\rm c}}
\;
\frac{\mathrm{d}}{\mathrm{d}\nu}\overline{E}^{a_{\rm x},\nu}
\;\mathrm{d}\nu,
\end{eqnarray}
could be used, but then the connection to $\lambda_1$-B2-PLYP would be
lost, simply because different intervals are used. In fact, segmenting the B2-PLYP energy in the same manner as the $\lambda_1$-B2-PLYP one is
not trivial. The main reason is that, in the $\lambda_1$-B2-PLYP scheme,
$\lambda_1$ and $\lambda_2$ are independent parameters but $a_{\rm c}$
and $a_{\rm x}$ are {\it not}, since the former depends on both
$\lambda_1$ and $\lambda_2$.
On the other hand, in B2-PLYP,    
$a_{\rm c}$ and $a_{\rm x}$ are independent parameters. Considering
that, in the particular case $a_{\rm c}=a^2_{\rm x}$ or equivalently
$\lambda_1=\lambda_2$, B2-PLYP and $\lambda_1$-B2-PLYP are identical
($\overline{E}^{\nu,\nu^2}=\tilde{E}^{\nu,\nu}$), we
propose the following segmentation, by analogy with
Eq.~(\ref{segments_lambda1b2plyp}), 
\begin{eqnarray}\label{convb2plyp_segmentation}
\overline{E}^{a_{\rm x},a_{\rm c}}&=&
\overline{E}^{0,0}
+
\displaystyle
\int^{a_{\rm x}-\sqrt{a^2_{\rm x}-a_{\rm c}}}_0
\;
\frac{\mathrm{d}}{\mathrm{d}\nu}\overline{E}^{\nu,\nu^2}
\mathrm{d}\nu
\nonumber\\
&&
\displaystyle
+\int^{a_{\rm x}}_{a_{\rm x}-\sqrt{a^2_{\rm x}-a_{\rm c}}}
\;
\frac{\mathrm{d}}{\mathrm{d}\nu}\Big(\overline{E}^{\nu,\nu^2}
-\overline{E}^{a_{\rm x},a^2_{\rm x}-(a_{\rm x}-\nu)^2}\Big)
\;\mathrm{d}\nu.
\end{eqnarray}
Note that the derivative $\mathrm{d}\overline{E}^{\nu,\nu^2}/\mathrm{d}\nu$ is integrated up to
$a_{\rm x}$, which ensures that the orbitals are calculated with the
fraction $a_{\rm x}$ of HF exchange. This is an important difference
with $\lambda_1$-B2-PLYP for which this fraction equals $\lambda_1=a_{\rm x}-\sqrt{a^2_{\rm
x}-a_{\rm c}}$ instead. As a result, the
corresponding fraction of MP2 correlation energy must be reduced from $a^2_{\rm x}$ to $a_{\rm c}$, which is exactly what the
third term in Eq.~(\ref{convb2plyp_segmentation}) is devoted to. 
Finally, since the particular case $a_{\rm x}=a_{\rm c}=0$ corresponds
to a standard BLYP calculation, the conventional B2-PLYP energy can be
rewritten, according Eqs.~(\ref{Eblyp}) and (\ref{convb2plyp_segmentation}) 
, and Appendix~\ref{Asec:B2PLYPInt}, in terms of an exchange--correlation
integrand 
\begin{eqnarray}\label{convb2plypintegrand}
\overline{E}^{a_{\rm x},a_{\rm c}}
&=& 
\langle\tilde{\Phi}^0 \vert
\hat{T}+\hat{V}\vert\tilde{\Phi}^0\rangle
+{E}_{\rm H}[n_{\tilde{\Phi}^0}]
+\displaystyle
\int^1_0
\mathcal{\overline{W}}^{a_{\rm x},a_{\rm c},\nu}_{\rm xc}
\mathrm{d}\nu
,
\end{eqnarray}
where the exchange part is defined, like in the $\lambda_1$-B2-PLYP
scheme, from the HF and Becke exchange energies calculated for the
KS-BLYP determinant 
\begin{eqnarray}\label{convb2lypXintegranddef}
\mathcal{\overline{W}}^{a_{\rm x},\nu}_{\rm x}&=&
E^{\mbox{\tiny HF}}_{\rm x}[\tilde{\Phi}^{0}]
\times
\mathcal{I}_{[0,a_{\rm x}[}(\nu)
+E^{\mbox{\tiny B}}_{\rm x}[n_{\tilde{\Phi}^0}]
	     \times \mathcal{I}_{[
	    a_{\rm x}, 1]}(\nu),
\end{eqnarray}
and the associated correlation integrand equals
\begin{eqnarray}\label{convb2plyp_Cintegrand}
\mathcal{\overline{W}}^{a_{\rm x},a_{\rm c},\nu}_{\rm c}
&=&
\mathcal{\overline{W}}^{a_{\rm x},a_{\rm c},\nu}_{\rm xc}
-
\mathcal{\overline{W}}^{a_{\rm x},\nu}_{\rm x}
\nonumber\\
&=&
\Bigg(
E^{\mbox{\tiny B}}_{\rm x}[n_{\tilde{\Phi}^0}]
-E^{\mbox{\tiny B}}_{\rm x}[n_{\tilde{\Phi}^{\nu}}]
+E^{\mbox{\tiny HF}}_{\rm x}[\tilde{\Phi}^{\nu}]
-E^{\mbox{\tiny HF}}_{\rm x}[\tilde{\Phi}^{0}]
\nonumber\\
&&+2\nu
\overline{E}^{(2)a_{\rm x},a_{\rm c}}_{\rm
\mbox{\tiny MP}}
+2\nu \big(E^{\mbox{\tiny
LYP}}_{\rm c}[n_{\tilde{\Phi}^0}]
-E^{\mbox{\tiny
LYP}}_{\rm
c}[n_{\tilde{\Phi}^{\nu}}]\big) 
\Bigg)
\nonumber     \\
&&
\times \mathcal{I}_{[0,a_{\rm x}-\sqrt{a^2_{\rm x}-a_{\rm c}}[}(\nu)\nonumber\\
&&+
\Bigg(
E^{\mbox{\tiny B}}_{\rm x}[n_{\tilde{\Phi}^0}]
-E^{\mbox{\tiny B}}_{\rm x}[n_{\tilde{\Phi}^{\nu}}]
+E^{\mbox{\tiny HF}}_{\rm x}[\tilde{\Phi}^{\nu}]
-E^{\mbox{\tiny HF}}_{\rm x}[\tilde{\Phi}^{0}]
\nonumber     \\
&&
+2\big(a_{\rm x}-\sqrt{a^2_{\rm x}-a_{\rm c}}\big)
\overline{E}^{(2)a_{\rm x},a_{\rm c}}_{\rm
\mbox{\tiny MP}}
\nonumber\\
&&
+2\nu \big(E^{\mbox{\tiny
LYP}}_{\rm c}[n_{\tilde{\Phi}^0}]
-E^{\mbox{\tiny
LYP}}_{\rm
c}[n_{\tilde{\Phi}^{\nu}}]\big)\nonumber\\
&&
+2(a_{\rm x}-\nu)
{E}^{\mbox{\tiny LYP}}_{\rm
c}[n_{\overline{\Phi}^{a_{\rm x},a^2_{\rm x}-(a_{\rm x}-\nu)^2}}]
\Bigg)
\nonumber\\
&&
 \times    \mathcal{I}_{[a_{\rm x}-\sqrt{a^2_{\rm x}-a_{\rm c}},a_{\rm
 x}[}(\nu)\nonumber\\
&&+
2\nu E^{\mbox{\tiny LYP}}_{\rm c}[n_{\tilde{\Phi}^0}]
	     \times \mathcal{I}_{[
	    a_{\rm x}, 1]}(\nu).
\end{eqnarray}
Comparing Eqs.~(\ref{lambda1b2plyp_Cintegrand}) and
(\ref{convb2plyp_Cintegrand}) it is clear that, in the first segment, the B2-PLYP
and $\lambda_1$-B2-PLYP correlation integrands are formally identical.
The only difference lies in the MP2 correlation energies, which are not
calculated with the same set of orbitals, as discussed previously. A qualitative behaviour of the correlation integrand along the adiabatic
connection is obtained when neglecting the variation of all terms that depend implicitly
on $\nu$:
\begin{eqnarray}\label{convb2plyp_qualtitativ_Cintegrand}
\mathcal{\overline{W}}^{a_{\rm x},a_{\rm c},\nu}_{\rm c}&\sim&
2\nu
\overline{E}^{(2)a_{\rm x},a_{\rm c}}_{\rm
\mbox{\tiny MP}}
\times
\mathcal{I}_{[0,a_{\rm x}-\sqrt{a^2_{\rm x}-a_{\rm c}}[}(\nu)\nonumber\\
&&
-2\nu
{E}^{\mbox{\tiny LYP}}_{\rm
c}[n_{\tilde{\Phi}^0}]
 \times    \mathcal{I}_{[a_{\rm x}-\sqrt{a^2_{\rm x}-a_{\rm c}},a_{\rm
 x}[}(\nu)\nonumber\\
&&+
2\nu E^{\mbox{\tiny LYP}}_{\rm c}[n_{\tilde{\Phi}^0}]
	     \times \mathcal{I}_{[
	    a_{\rm x}, 1]}(\nu).
\end{eqnarray}
A striking difference with the $\lambda_1$-B2-PLYP correlation integrand
(see Eq.~(\ref{lambda1b2plyp_qualitativeCintegrand})) is the positive
slope in the second segment. This unphysical behaviour is
directly related to our definition of the B2-PLYP integrand.
It is a simple illustration of the
fact that, when $a_{\rm c}<a^2_{\rm x}$, B2-PLYP does not rely on the density-functional perturbation
theory we derived, by contrast to $\lambda_1$-B2-PLYP. Still, after
integration over the second segment, the B2-PLYP integrand provides an
energy contribution which differs from the $\lambda_1$-B2-PLYP one as 
\begin{eqnarray}\label{comp_lambda1_int_2ndsegment}
&&
\displaystyle
\int_{a_{\rm x}-\sqrt{a^2_{\rm x}-a_{\rm c}}}^{a_{\rm x}}
\;
\mathcal{\overline{W}}^{a_{\rm x},a_{\rm c},\nu}_{\rm
c}-\mathcal{\tilde{W}}^{a_{\rm x}-\sqrt{a^2_{\rm x}-a_{\rm c}},a_{\rm x},\nu}_{\rm c}
\mathrm{d}\nu
\nonumber\\
&\approx&
\displaystyle
\int_{a_{\rm x}-\sqrt{a^2_{\rm x}-a_{\rm c}}}^{a_{\rm x}}
\;
2\big(a_{\rm x}-\sqrt{a^2_{\rm x}-a_{\rm
c}}\big)\Big(\overline{E}^{(2)a_{\rm x},a_{\rm c}}_{\rm \mbox{\tiny MP}}
-\tilde{E}^{(2)a_{\rm x}-\sqrt{a^2_{\rm x}-a_{\rm c}}}_{\mbox{\tiny MP}}\Big)
\mathrm{d}\nu
\nonumber\\
&&+\displaystyle
\int_{a_{\rm x}-\sqrt{a^2_{\rm x}-a_{\rm c}}}^{a_{\rm x}}
\;2\Big(2a_{\rm x}-\sqrt{a^2_{\rm x}-a_{\rm c}}-2\nu)
\Big)
{E}^{\mbox{\tiny LYP}}_{\rm
c}[n_{\tilde{\Phi}^0}]
\nonumber\\
&=&
2\big(a_{\rm x}-\sqrt{a^2_{\rm x}-a_{\rm
c}}\big)\sqrt{a^2_{\rm x}-a_{\rm c}}\times\Big(\overline{E}^{(2)a_{\rm x},a_{\rm c}}_{\rm \mbox{\tiny MP}}
-\tilde{E}^{(2)a_{\rm x}-\sqrt{a^2_{\rm x}-a_{\rm c}}}_{\mbox{\tiny
MP}}\Big),
\end{eqnarray}
if we neglect the variation of all terms that depend implicitly on
$\nu$. As a result, B2-PLYP and $\lambda_1$-B2-PLYP correlation energies will
essentially differ by the MP2 term. 

Finally, we remark that since the B2-PLYP energy was not derived by consideration of the AC directly it is possible to construct a number of AC integrands for this approach. One alternative segmentation has already been presented in Eq.~(\ref{convb2plyp_othersegmentation}). However, another possibility is to regard the B2-PLYP parameters as entirely empirical parameters, which simply scale the ACs derived for each component by a constant at all values of the interaction strength. A smooth AC integrand for B2-PLYP can then be obtained by summing these scaled components. However, whilst this integrand can be compared with the \textit{ab initio} curves, the connection to the density-functional perturbation theory and the $\lambda_1$-B2-PLYP methods presented here is lost.   

\subsection{Summary}\label{subsec:theory:summary}


A three-part segmentation of the exact exchange--correlation integrand has been proposed, which is directly connected to the double hybrid functionals of Ref.~\cite{2blehybrids_Fromager2011JCP}. 
Each segment of the AC has been expanded through second order
within density-functional perturbation theory. When neglecting both
second-order corrections to the density and density scaling, and using
the Becke exchange functional in conjunction with the LYP correlation
functional, the $\lambda_1$-B2-PLYP  integrand is obtained. An integrand
expression has also been derived for the conventional B2-PLYP scheme.
Both schemes are completely equivalent when $a^2_{\rm x}=a_{\rm c}$ or,
equivalently, $\lambda_1=\lambda_2$. In
this case, the second segment simply disappears.
Interestingly for standard $a_{\rm x}=0.53$ and $a_{\rm c}=0.27$
values~\cite{2blehybrids_Grimme}, $a^2_{\rm x}\approx 0.28$ differs only by 0.01 from $a_{\rm c}$,
as already pointed out by Sharkas {\it et
al.}~\cite{2blehybrids_Julien}. Still, since $\lambda_2$ differs from
$\lambda_1$ by $\sqrt{a^2_{\rm x}-a_{\rm c}}\approx 0.1$ (see Eq.~(\ref{axac})),
the second segment represents $10\%$ of the total AC which is not
negligible. 

For comparison the methods used to determine accurate \textit{ab initio} and pure density-functional estimates of the integrands have been outlined. All the schemes investigated in this work are summarized in Table~\ref{tab:summary_integrands}.
\begin{table}
\caption{The exchange--correlation integrands computed in this
work ($\tilde{\Phi}^0$ denotes the KS-BLYP determinant).}\label{tab:summary_integrands}
\centering
\begin{tabular}{lccc}
\hline
\hline
 {Integrand} & \text{Exchange} & \text{Correlation} &
 \text{Parameters} \\
 \hline
 {\it ab initio}& Eq.~(\ref{abinitio_EXX})&
 Eq.~(\ref{WACCI}) & -\\
 $\lambda_1$-B2-PLYP & Eq.~(\ref{lambda1b2lypXCintegranddef2}) &
 Eq.~(\ref{lambda1b2plyp_Cintegrand}) & 
 $\lambda_1\approx 0.426^a$,
 $\lambda_2=0.53^a$
 \\
 $\lambda_1$-DS-B2-PLYP & Eq.~(\ref{lambda1b2lypXCintegranddef2}) &
 Eq.~(\ref{DSlambda1b2plyp_Cintegrand}) &
 $\lambda_1\approx 0.426$,
 $\lambda_2=0.53$
 \\
 B2-PLYP &Eq.~(\ref{convb2lypXintegranddef}) &
 Eq.~(\ref{convb2plyp_Cintegrand}) & 
 $a_{\rm x}=0.53^b$, $a_{\rm c}=0.27^b$
 \\
 BLYP & $E^{\mbox{\tiny B}}_{\rm x}[n_{\tilde{\Phi}^0}]$ &$\Delta^{\mbox{\tiny
 LYP},\nu}_{\rm c}[n_{\tilde{\Phi}^0}]^c$
& -\\
\hline
\hline
\end{tabular}
\newline
\flushleft
\hspace{0.2in}$^a$ see Eq.~(\ref{axac})\\
\hspace{0.2in}$^b$ Ref.~\cite{2blehybrids_Grimme}\\
\hspace{0.2in}$^c$ see Eq.~(\ref{deltaclambdaexpr_scaled_density_blyp})
\end{table}

\section{Results and discussion}\label{sec:results}

In the present work we study the AC integrands for the species, H$_2$,
(He)$_2$, He-Ne, LiH, HF, N$_2$ and H$_2$O. For H$_2$ we consider the
geometries $R=1.4$ and $3.0$ a.u., for the (He)$_2$ and He-Ne van der
Waals dimers we perform calculations at the equilibrium geometries of
$5.612$ a.u.~\cite{Ogilvie:1992} and $5.728$ a.u.~\cite{Ogilvie:1993},
respectively. For the remaining four molecular systems LiH, HF, N$_2$
and H$_2$O we use the equilibrium geometries of Ref.~\cite{Teale:2013}
calculated at the CCSD(T)/cc-pVTZ level with all electrons correlated.
All calculations of the AC integrands are performed in
the aug-cc-pVTZ basis
set~\cite{DUNNING:1989p1833,KENDALL:1992p1831,WOON:1994p1834} using a
modified version of the DALTON2011 program~\cite{DALTON}, which contains
implementations of the methodologies outlined in
Section~\ref{sec:CompAC} and summarized in
Table~\ref{tab:summary_integrands}.    

\subsection{\textit{ab initio} ACs}
We have performed calculations using the methodology described in Section~\ref{subsec:theory:approxac:abinit} for the species above. A range of interaction strengths in the interval $\nu \in [0,1]$, have been considered. In order to account for rapid curvature in the low-$\nu$ part of the curve characteristic of statically correlated systems we have used the $\nu$ values $\{0, 10^{-6}, 10^{-5}, 10^{-4}, 10^{-3}, 10^{-2}, 0.1, 0.2, 0.3, 0.4, 0.5, 0.6, 0.7, 0.8, 0.9, 1.0\}$. In Ref.~\cite{Teale:2010} a form for the AC integrand was proposed based on consideration of a simple two-state CI model and shown to have sufficient flexibility to reproduce correlation energies in systems exhibiting both static and dynamical correlation effects. Here we perform a least squares fit of this form
\begin{eqnarray}
\mathcal{W}_\text{c}^{\nu,\text{AC-CI}} &=& - \frac{1+\sqrt{5}}{4}a 
\nonumber\\
&&
-
\frac{4(2+\sqrt{5})a^2 +5(3+\sqrt{5})as\nu}{2 \sqrt{ 8(7+3\sqrt{5})a^2 +
16(2+\sqrt{5})as\nu + 10(3+ \sqrt{5})s^2 \nu^2}}, \label{WACCI}
\end{eqnarray}  
to the calculated \textit{ab initio} estimates of $\mathcal{W}_{\rm
c}^\nu[n]$ at
the values of $\nu$ above. The fitted values of the parameters $a$ and
$s$ are reported in Table~\ref{tab:fitcoeffs} for each species in this
study. Also reported is the quantity $\Delta E_{\text{c}} = \int_0^1
\mathcal{W}_\text{c}^{\nu,\text{AC-CI}} \mathrm{d}\nu - 
(
E_{\rm CC}^{\rm tot.} - E_{\rm nn} - T_{\rm s}[n] - E_{\rm ne}[n] - E_{\text{H}}[n] - E_\text{x}[n])$
which provides a consistency check for the quality of the fitted function as compared with the explicitly calculated correlation energy using non-interacting and interacting energies. As can be seen in Table~\ref{tab:fitcoeffs} these values are reasonable and of sufficient accuracy to allow these fitted functions to serve as a benchmark against which to compare double-hybrid integrands.
\begin{table}
\caption{Fitted coefficients $a$ and $s$ in Eq.~(\ref{WACCI}) for the species considered in this work. Also shown are the correlation energies calculated by integration of the fitted curves and the difference between these values and those calculated as $\Delta E_\text{c} = \int_0^1 \mathcal{W}_\text{c}^{\nu, \text{AC-CI}}\mathrm{d}\nu - (E_{\rm CC}^{\rm tot.} - E_{\rm nn} - T_{\rm s}[n] - E_{\rm ne}[n] - E_{\text{H}}[n] - E_\text{x}[n])$}\label{tab:fitcoeffs}
\centering
\begin{tabular}{lrrrr}
\hline
\hline
 \text{Molecule} & $a$ & $s$ & $E_\text{c}$ & $\Delta E_\text{c}$ \\
 \hline
 \text{H$_2$ (R=1.4 a.u.)} & \text{ $-$0.171004} & \text{ $-$0.095425} & \text{ $-$0.039851} &$ 6.91\times 10^{-6}$ \\
 \text{H$_2$ (R=3.0 a.u.)} & \text{ $-$0.153978} & \text{ $-$0.255931} & \text{ $-$0.076559} & $-3.82\times 10^{-5}$ \\
 \text{(He)$_2$} & \text{ $-$0.513334} & \text{ $-$0.176830} & \text{ $-$0.079183} & $5.62\times 10^{-6}$ \\
 \text{He-Ne} & \text{ $-$1.393157} & \text{ $-$0.806875} & \text{ $-$0.334652} & $4.38\times 10^{-4}$ \\
 \text{HF} & \text{ $-$1.061605} & \text{ $-$0.775661} & \text{ $-$0.306331} & $8.03\times 10^{-4}$ \\
 \text{LiH} & \text{ $-$0.220273} & \text{ $-$0.128740} & \text{ $-$0.053303} & $6.37\times 10^{-5}$ \\
 \text{N$_2$} & \text{ $-$1.226312} & \text{ $-$1.201367} & \text{ $-$0.438569} & $1.56\times 10^{-3}$ \\
 \text{H$_2$O} & \text{ $-$0.993788} & \text{ $-$0.775566} & \text{ $-$0.301455} & $8.37\times 10^{-4}$ \\
 \hline
 \hline
\end{tabular}
\end{table}

\subsection{The H$_2$ molecule}\label{subsec:results:H2}

The H$_2$ molecule has been widely studied as a prototypical system, see e.g. Refs.~\cite{BUIJSE:1989p1838,BAERENDS:2001p1817}, exhibiting a smooth transition from predominantly dynamical correlation effects at short and equilibrium $R$ values to predominantly static correlation effects a large $R$ values. It has been argued by Gritsenko \emph{et al.}~\cite{GRITSENKO:1996p1796} that for single hybrid exchange-based functionals the optimal fraction of orbital dependent exchange varies with $R$, approaching zero as $R$ increases. More recently in Ref.~\cite{Teale:2010} an analysis of the BLYP AC integrand, in comparison with the FCI AC integrand, showed that close to equilibrium $R$ the BLYP functional provides a reasonable estimate of exchange and correlation energies. However for larger $R$ values beyond $\sim 7$ a.u. the estimate of the exchange energy is significantly too negative, whilst the correlation energy is significantly too positive. This is manifested by a much too flat shape for the BLYP correlation AC integrand at these geometries. At intermediate geometries $R \approx 3$ a.u. error cancellations between the exchange and correlation energies can lead to reasonable total energies.

The exchange energy contributions to the two-parameter double hybrids
are shown in Table~\ref{tab:Ex}. The individual HF and
density-functional type contributions are shown, calculated on the
KS-BLYP $\tilde{\Phi}^0$ determinant and $n_{\tilde{\Phi}^0}$ density,
respectively. Their weighted contribution, of relevance to the
double-hybrid functionals, is also tabulated. For comparison the
exchange energies relevant to the \textit{ab initio} estimates of the AC
are also included, these are evaluated from the KS orbitals at $\nu=0$
which give the FCI density. For the H$_2$ molecule at $R=1.4$ a.u. it is
clear that both the HF and density-functional estimates of the exchange
energy are comparable. Their weighted average is also close to the
exchange energy calculated for the KS orbitals giving the FCI density.
At the longer bond length of $R=3.0$ a.u. the difference between the HF and density-functional exchange contributions is much more pronounced. Comparing with the accurate FCI value in the same basis set it is clear that the density-functional gives a much too negative exchange energy, as was also noted in Ref.~\cite{Teale:2010}. Here we see that the weighted average used in the double-hybrid approaches significantly reduces this error. 
\begin{table}
\caption{Exchange energy contributions for the $\lambda_1$-B2-PLYP and
B2-PLYP double hybrid schemes. The reference determinant
$\tilde{\Phi}^0$ is the KS-BLYP one. The HF exchange weight is set to   
$a_\text{x}=0.53$.
For comparison the exchange energies $E^{\rm HF}_{\rm
x}[\Phi^{\text{KS}}]$ of KS determinants constrained to yield accurate \textit{ab initio} densities have also been included (see text for details). All values are given in atomic units.}\label{tab:Ex}
\begin{tabular}{lrrrr}
\hline
\hline
&$E^{\mbox{\tiny HF}}_{\rm x}[\tilde{\Phi}^{0}]$&$E^{\mbox{\tiny B}}_{\rm
x}[n_{\tilde{\Phi}^0}]$&$a_{\rm x}\,E^{\mbox{\tiny HF}}_{\rm
x}[\tilde{\Phi}^{0}]$&$E^{\mbox{\tiny HF}}_{\rm x}
[\Phi^{\text{KS}}]
$\\
&   &   & $+(1-a_{\rm x})\,E^{\mbox{\tiny B}}_{\rm x}[n_{\tilde{\Phi}^0}]$ &   \\
\hline 
H$_2$ ($R=1.4$ a.u.) & $-$0.6566   & $-$0.6563    & $-$0.6565 &  $-$0.6608 \\
H$_2$ ($R=3.0$ a.u.) & $-$0.4720  & $-$0.5061  &   $-$0.4880  &   $-$0.4769  \\
He$_2$   &  $-$2.0295   &  $-$2.0364    &   $-$2.0327    &   $-$2.0460  \\
HeNe       &  $-$13.0517  &  $-$13.1084   &  $-$13.0783  & $-$13.0861\\
LiH        &    $-$2.1297         &   $-$2.1292        &    $-$2.1294      &  $-$2.1369\\
HF        &     $-$10.3709      &    $-$10.4404      &   $-$10.4036    &  $-$10.3870\\
N$_2$  &     $-$13.0855      &    $-$13.1977      &   $-$13.1382    &  $-$13.0888\\
H$_2$O&    $-$8.9062        &    $-$8.9674        &    $-$8.9350     &  $-$8.9149 \\
\hline
\hline
\end{tabular}\\
\end{table}

\begin{figure}
\centering
{\includegraphics[scale=1]{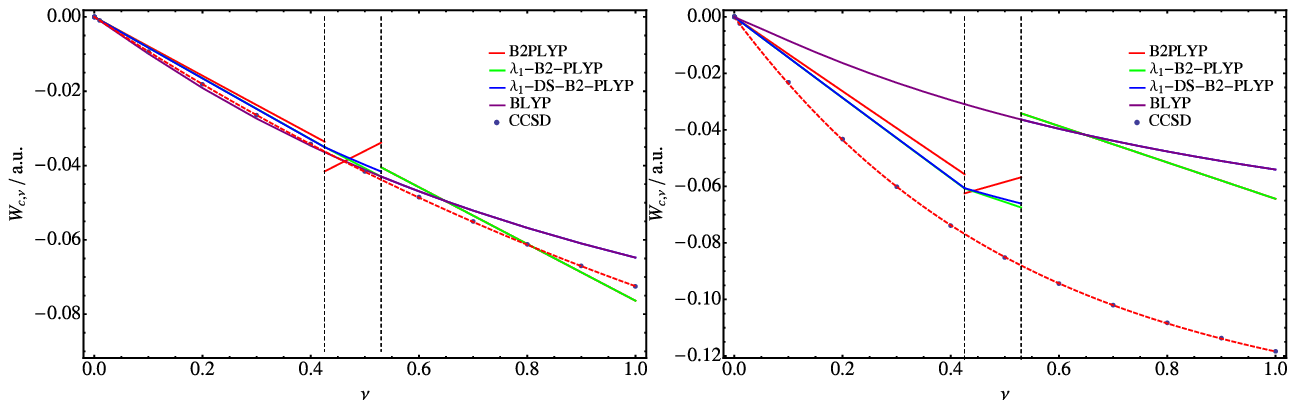}}
\caption{ACs for the H$_2$ molecule at $R=1.4$ a.u.
(left panel) and $R=3.0$ a.u. (right panel) calculated using the
conventional B2-PLYP functional (solid red line), the
$\lambda_1$-B2-PLYP two parameter double hybrid (green line), the
$\lambda_1$-DS-B2-PLYP functional (blue line) (which includes coordinate scaling contributions) and the standard BLYP functional (purple line). Also included is an accurate \textit{ab initio} AC calculated at the FCI level (blue points). The function in Eq.~(\ref{WACCI}) has been fitted to this data (red dashed line) and the coefficients for this form are reported in Table~\ref{tab:fitcoeffs}.}\label{fig:H2}
\end{figure}

In Fig.~\ref{fig:H2} we present the correlation integrands for the double-hybrid approximations, as well as the BLYP and \textit{ab initio} estimates. In the left-hand panel the correlation AC integrands for the methods considered are shown at $R=1.4$ a.u. As was shown in Ref.~\cite{Teale:2010} at this geometry the BLYP AC integrand is reasonable, though it tends to be too positive for larger $\nu$ values. The challenge for double-hybrid approaches is to provide a model AC integrand which improves over the pure DFT integrand (in this case BLYP) whilst utilizing the DFT integrand where it is accurate. For the $R=1.4$ a.u. geometry the total correlation energies in Table~\ref{tab:corEAC} are all quite similar and close to the FCI estimate. This is also clear graphically from Fig.~\ref{fig:H2}. 
\begin{table}
\caption{The correlation energies for each segment of the double-hybrid
ACs.
For comparison the values for the correlation energies calculated from
BLYP as well as accurate \textit{ab initio} estimates of the AC 
are included.}\label{tab:corEAC}
\centering
\begin{tabular}{llrrrr}
\hline
\hline
 \text{Molecule} & \text{Method} & $E_\text{c}^{\text{Seg 1}}$ & $E_\text{c}^{\text{Seg 2}}$ & $E_\text{c}^{\text{Seg 3}}$ & $E_\text{c}^{\text{Tot.}}$ \\
\hline
\text{H$_2$ ($R=1.4$ a.u.)} & \text{B2-PLYP} & \text{   $-$0.0072} & \text{   $-$0.0039} & \text{   $-$0.0275} & \text{   $-$0.0385} \\
 \text{} & \text{$\lambda_1$-B2-PLYP} & \text{   $-$0.0075} & \text{   $-$0.0038} & \text{   $-$0.0275} & \text{   $-$0.0387} \\
 \text{} & \text{$\lambda_1$-DS-B2-PLYP} & \text{   $-$0.0075} & \text{   $-$0.0040} & \text{   $-$0.0257} & \text{   $-$0.0372} \\
 \text{} & \text{BLYP} & \text{   $-$0.0083} & \text{   $-$0.0041} & \text{   $-$0.0257} & \text{   $-$0.0382} \\
 \text{} & \text{FCI} & \text{   $-$0.0080} & \text{   $-$0.0042} & \text{   $-$0.0276} & \text{   $-$0.0399} \\
 \hline
 \text{H$_2$ ($R=3.0$ a.u.)} & \text{B2-PLYP} & \text{   $-$0.0119} & \text{   $-$0.0062} & \text{   $-$0.0231} & \text{   $-$0.0413} \\
 \text{} & \text{$\lambda_1$-B2-PLYP} & \text{   $-$0.0130} & \text{   $-$0.0067} & \text{   $-$0.0231} & \text{   $-$0.0428} \\
 \text{} & \text{$\lambda_1$-DS-B2-PLYP} & \text{   $-$0.0129} & \text{   $-$0.0066} & \text{   $-$0.0216} & \text{   $-$0.0412} \\
 \text{} & \text{BLYP} & \text{   $-$0.0071} & \text{   $-$0.0035} & \text{   $-$0.0216} & \text{   $-$0.0322} \\
 \text{} & \text{FCI} & \text{   $-$0.0184} & \text{   $-$0.0086} & \text{   $-$0.0495} & \text{   $-$0.0765} \\
 \hline
 \text{(He)$_2$} & \text{B2-PLYP} & \text{   $-$0.0146} & \text{   $-$0.0081} & \text{   $-$0.0630} & \text{   $-$0.0857} \\
 \text{} & \text{$\lambda_1$-B2-PLYP} & \text{   $-$0.0150} & \text{   $-$0.0083} & \text{   $-$0.0630} & \text{   $-$0.0863} \\
 \text{} & \text{$\lambda_1$-DS-B2-PLYP} & \text{   $-$0.0151} & \text{   $-$0.0082} & \text{   $-$0.0598} & \text{   $-$0.0830} \\
 \text{} & \text{BLYP} & \text{   $-$0.0184} & \text{   $-$0.0094} & \text{   $-$0.0598} & \text{   $-$0.0876} \\
 \text{} & \text{CCSD(T)} & \text{   $-$0.0153} & \text{   $-$0.0081} & \text{   $-$0.0557} & \text{   $-$0.0792} \\
 \hline
 \text{HeNe} & \text{B2-PLYP} & \text{   $-$0.0685} & \text{   $-$0.0384} & \text{   $-$0.3070} & \text{   $-$0.4138} \\
 \text{} & \text{$\lambda_1$-B2-PLYP} & \text{   $-$0.0706} & \text{   $-$0.0392} & \text{   $-$0.3070} & \text{   $-$0.4168} \\
 \text{} & \text{$\lambda_1$-DS-B2-PLYP} & \text{   $-$0.0707} & \text{   $-$0.0386} & \text{   $-$0.2899} & \text{   $-$0.3992} \\
 \text{} & \text{BLYP} & \text{   $-$0.0913} & \text{   $-$0.0457} & \text{   $-$0.2900} & \text{   $-$0.4270} \\
 \text{} & \text{CCSD(T)} & \text{   $-$0.0678} & \text{   $-$0.0347} & \text{   $-$0.2267} & \text{   $-$0.3292} \\
 \hline
 \text{LiH} & \text{B2-PLYP} & \text{   $-$0.0098} & \text{   $-$0.0058} & \text{   $-$0.0637} & \text{   $-$0.0792} \\
 \text{} & \text{$\lambda_1$-B2-PLYP} & \text{   $-$0.0102} & \text{   $-$0.0060} & \text{   $-$0.0637} & \text{   $-$0.0799} \\
 \text{} & \text{$\lambda_1$-DS-B2-PLYP} & \text{   $-$0.0103} & \text{   $-$0.0057} & \text{   $-$0.0586} & \text{   $-$0.0746} \\
 \text{} & \text{BLYP} & \text{   $-$0.0204} & \text{   $-$0.0096} & \text{   $-$0.0586} & \text{   $-$0.0886} \\
 \text{} & \text{CCSD(T)} & \text{   $-$0.0108} & \text{   $-$0.0056} & \text{   $-$0.0368} & \text{   $-$0.0532} \\
 \hline
 \text{HF} & \text{B2-PLYP} & \text{   $-$0.0646} & \text{   $-$0.0357} & \text{   $-$0.2602} & \text{   $-$0.3606} \\
 \text{} & \text{$\lambda_1$-B2-PLYP} & \text{   $-$0.0676} & \text{   $-$0.0371} & \text{   $-$0.2602} & \text{   $-$0.3650} \\
 \text{} & \text{$\lambda_1$-DS-B2-PLYP} & \text{   $-$0.0677} & \text{   $-$0.0364} & \text{   $-$0.2438} & \text{   $-$0.3480} \\
 \text{} & \text{BLYP} & \text{   $-$0.0791} & \text{   $-$0.0390} & \text{   $-$0.2439} & \text{   $-$0.3620} \\
 \text{} & \text{CCSD(T)} & \text{   $-$0.0638} & \text{   $-$0.0319} & \text{   $-$0.2027} & \text{   $-$0.2984} \\
 \hline
 \text{N$_2$} & \text{B2-PLYP} & \text{   $-$0.0952} & \text{   $-$0.0522} & \text{   $-$0.3480} & \text{   $-$0.4954} \\
 \text{} & \text{$\lambda_1$-B2-PLYP} & \text{   $-$0.1002} & \text{   $-$0.0544} & \text{   $-$0.3480} & \text{   $-$0.5026} \\
 \text{} & \text{$\lambda_1$-DS-B2-PLYP} & \text{   $-$0.1002} & \text{   $-$0.0534} & \text{   $-$0.3225} & \text{   $-$0.4762} \\
 \text{} & \text{BLYP} & \text{   $-$0.1088} & \text{   $-$0.0526} & \text{   $-$0.3226} & \text{   $-$0.4840} \\
 \text{} & \text{CCSD(T)} & \text{   $-$0.0954} & \text{   $-$0.0455} & \text{   $-$0.2771} & \text{   $-$0.4180} \\
 \hline
 \text{H$_2$O} & \text{B2-PLYP} & \text{   $-$0.0632} & \text{   $-$0.0348} & \text{   $-$0.2445} & \text{   $-$0.3425} \\
 \text{} & \text{$\lambda_1$-B2-PLYP} & \text{   $-$0.0658} & \text{   $-$0.0360} & \text{   $-$0.2445} & \text{   $-$0.3463} \\
 \text{} & \text{$\lambda_1$-DS-B2-PLYP} & \text{   $-$0.0658} & \text{   $-$0.0352} & \text{   $-$0.2272} & \text{   $-$0.3282} \\
 \text{} & \text{BLYP} & \text{   $-$0.0760} & \text{   $-$0.0369} & \text{   $-$0.2272} & \text{   $-$0.3402} \\
 \text{} & \text{CCSD(T)} & \text{   $-$0.0633} & \text{   $-$0.0314} & \text{   $-$0.1977} & \text{   $-$0.2924} \\
 \hline
 \hline
\end{tabular}
\end{table}

A number of differences between the methods become apparent when
examining the correlation integrand models graphically. For B2-PLYP we
see that the integrand in each segment is linear in the interaction
strength. For the first segment this is because the integrand is
dominated by a PT2 contribution based on a fixed set of orbitals. For
the third segment this is because the uniform coordinate scaling
contributions to the DFT correlation component are neglected. The most
striking feature however is the intermediate interval where the B2-PLYP
integrand is linear with positive slope and discontinuous at both
$\lambda_1$ and $\lambda_2$. The significance of this section and the ambiguity in the
choice of B2-PLYP AC were discussed in
Sec.~\ref{subsec:theory:approxac:convB2PLYP}. The $\lambda_1$-B2-PLYP
variant also has an integrand consisting of three linear segments,
however, continuity is restored at $\lambda_1$, although a derivative
discontinuity remains and a discontinuity is still present at
$\lambda_2$. Note that by definition the $\lambda_1$-B2-PLYP and B2-PLYP integrands are
identical in the third segment. The crossing in
the middle of the second segment, which implies that the 
$\lambda_1$-B2-PLYP and B2-PLYP correlation energies obtained by integration are very close (as
confirmed in Table~\ref{tab:corEAC}), is consistent with the small   
difference between the AC lines in the first segment (see
Sec.~\ref{subsec:theory:approxac:convB2PLYP}). The $\lambda_1$-DS-B2-PLYP variant
includes the effects due to uniform coordinate scaling. This affects the
slope in the second segment and makes the integrand coincide with the
BLYP integrand in the third segment. Interestingly, taking into account density
scaling in the LYP correlation functional does not improve, in this
particular case, the correlation energy when compared to FCI (see also
Table~\ref{tab:corEAC}).

In the right hand panel of Fig.~\ref{fig:H2} the same integrands are
presented for H$_2$ at $R=3.0$ a.u. Here the behaviour of the integrands
within each segment remains qualitatively similar, however, all of the
models now give a significantly too positive correlation energy as can
be seen from Table~\ref{tab:corEAC}. The discontinuous behaviour at
$\lambda_2$ also becomes much more pronounced. This can be understood by
noting that (as shown previously in Ref.~\cite{Teale:2010}) the BLYP
integrand is especially poor as the bond is stretched and static
correlation becomes more important. It is clear therefore that any
double-hybrid wishing to perform well for systems in which static
correlation plays a significant role must be constructed in a different
manner, no partitioning of the AC involving a pure density functional
component would seem to be advantageous. Furthermore, the neglect of
density scaling amounts to a linear approximation of the integrand and
so may affect the accuracy of the model even in regimes where dynamical
correlation is dominant. Note also, in the first segment, the  
difference between the $\lambda_1$-B2-PLYP and B2-PLYP slopes
which originates from the fact that the corresponding MP2 correlation energies are
calculated with different orbitals. These are mainly characterized by
the fraction of HF exchange which is larger for B2-PLYP (0.53) than for
$\lambda_1$-B2-PLYP (0.426). The difference becomes substantial upon
bond stretching. As expected from
Sec.~\ref{subsec:theory:approxac:convB2PLYP} the correlation energy
in the second segment is then larger (in absolute value) for
$\lambda_1$-B2-PLYP than for B2-PLYP. This is graphically illustrated in
the right panel of
Fig.~\ref{fig:H2} where the corresponding AC lines do not cross in the
middle of the segment, like in the equilibrium geometry (left panel).  

Since the double hybrids considered are based on PT2 theories one of course should not expect their range of applicability to extend to strongly correlated systems. Nonetheless these considerations may be helpful if one wishes to design a double-hybrid functional capable of describing molecules over a reasonable part of their potential energy surfaces around the equilibrium structure. Furthermore, it highlights the point that for overall improvement of double-hybrid approaches one should carefully consider not only the nature of the density functional approximations employed but also the wave-function contributions. 

As is discussed in Appendix~\ref{Asec:PTexpSegAC} we have chosen in the
present work to evaluate the PT2 contributions for the approximate
functionals on a fixed set of orbitals. As a consequence the
corresponding contributions to the AC are linear in the interaction
strength. If the orbital relaxation is taken into account at each
interaction strength then the same integrated value of the energy would
be obtained but the PT2 integrand would become curved due to a
dependence of the orbitals on the interaction strength (see Ref~\cite{Teale:2010} for further discussion). To obtain higher accuracy from the wave-function contribution to the double-hybrid functionals it would be desirable to introduce terms of higher order in the interaction strength without introducing significant extra computational cost. In this respect it is interesting to consider alternatives to PT2. Natural choices here would be higher order perturbation approaches or coupled-cluster type methodologies. However, the computational cost of these approaches is sufficiently high as to make them undesirable for application in this context. One interesting set of alternatives, which can be evaluated at a cost similar to that of PT2 theory and, as discussed by Furche~\cite{Furche:2001}, do contain higher order contributions in $\nu$ are the RPA correlation energies. Investigation of these variants of the correlation energy in the context of double-hybrid approaches based on a linear AC may be worthwhile. A number of empirical and range-separated approaches to combine DFT and RPA have already appeared in the literature, see for example, Refs.~\cite{Yan:2000,Furche:2001,rpa-srdft_toulouse,rpa-srdft_scuseria,Gruneis:2009,Eshuis:2010,Hesselmann:2010}.       

\subsection{(He)$_2$ and He-Ne van der Waals dimers}\label{subsec:results:vdW}

\begin{figure}
\centering
{\includegraphics[scale=1]{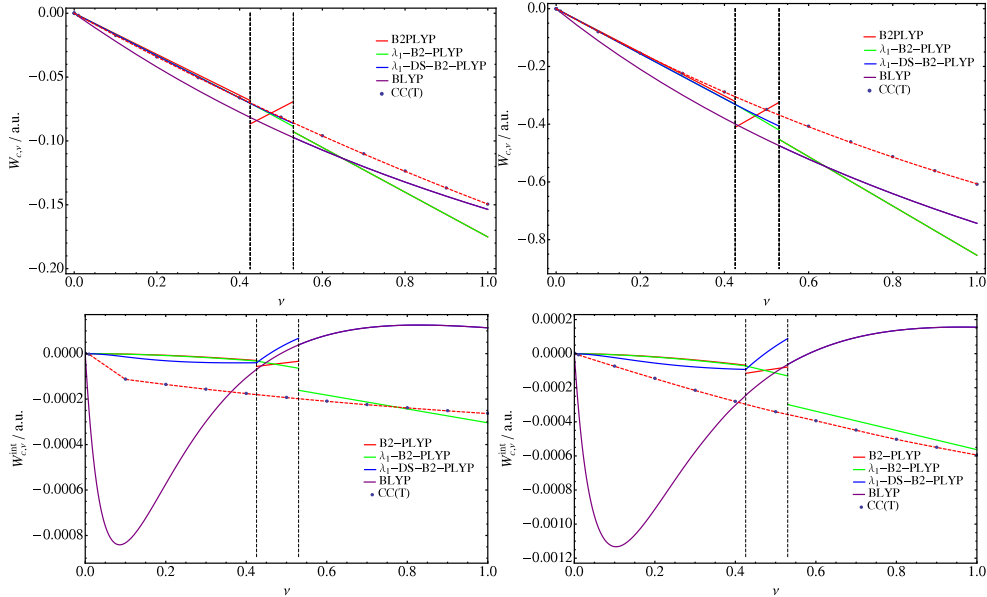}}
\caption{ACs for the van der Waals dimers (He)$_2$
(upper left panel) and He-Ne (upper right panel) calculated using the
conventional B2-PLYP functional (solid red line), the
$\lambda_1$-B2-PLYP two parameter double hybrid (green line), the
$\lambda_1$-DS-B2-PLYP functional (blue line) (which includes coordinate scaling contributions) and the standard BLYP functional (purple line). Also included is an accurate \textit{ab initio} estimate of the AC calculated at the CCSD(T) level (blue points). The function in Eq.~(\ref{WACCI}) has been fitted to this data (red dashed line) and the coefficients for this form are reported in Table~\ref{tab:fitcoeffs}. The lower panels ((He)$_2$ left and HeNe right) show the interaction ACs for each method as defined in Eq.~(\ref{Eq:IntAC}).}\label{fig:vdW}
\end{figure}

In Fig.~\ref{fig:vdW} we present the correlation integrands for two van der Waals dimers at their equilibrium geometries. The (He)$_2$ dimer has been widely studied as a prototypical system for examining van der Waals and dispersion interactions in DFT, see e.g. Refs.~\cite{allen:11113,Teale:2010b} and references therein. Methods which mix DFT with PT2 theory in a range-separated manner based on non-linear ACs have proven useful for the treatment of van der Waals and dispersion interaction energies. That the range separation of these interaction energies can be successful has recently been demonstrated by calculating \textit{ab initio} estimates of the AC integrands along non-linear paths. However, for conventional double hybrid approaches such as B2-PLYP the description of dispersion interactions is far less satisfactory. Indeed empirical dispersion corrections have been developed to add to this functional~\cite{Schwabe:2007}, despite its PT2 component.

The exchange energy contributions for the (He)$_2$ and He-Ne van der
Waals dimer systems are shown in Table~\ref{tab:Ex}. For the (He)$_2$
dimer the HF and density-functional estimates of the exchange energy are
similar and slightly more positive than the estimate based on KS
orbitals giving the CCSD(T) density, which we will denote KS[CCSD(T)].
The weighted average of the exchange energy relevant to the double
hybrids is therefore also reasonable. For the He-Ne dimer the HF estimate of the exchange energy is more positive than the KS[CCSD(T)] estimate, whilst the density-functional estimate is more negative. The weighted average is therefore much closer to the accurate value.

The upper two panels in Fig.~\ref{fig:vdW} show the total correlation
ACs for (He)$_2$ and He-Ne respectively. The shape of these curves are
similar in many respects to those observed for H$_2$ near its
equilibrium geometry. Although the standard BLYP functional now gives a
too negative integrand at all interaction strengths. The general
similarity between the (He)$_2$, He-Ne and H$_2$ ($R=1.4$ a.u.) curves
reflects the fact that on-atom dynamical correlation dominates the
overall correlation energy contribution. Still, by contrast with H$_2$,
density scaling improves the correlation energy of both van der Waals
dimers when compared to CCSD(T). These graphical results echo the observation by Sharkas {\it et al.}~\cite{2blehybrids_Julien} when computing atomization energies with
various double hybrid density-functionals; introducing density-scaling effects does not
systematically provide more accurate results when it is applied to
approximate correlation density-functionals. This clear in the present work when comparing the left hand panel of Fig.~\ref{fig:H2} and and upper left panel of Fig.~\ref{fig:vdW}. 

To examine the important correlation energy contributions to the interaction energies of the van der Waals dimers we have calculated the interaction ACs as in Ref.~\cite{Teale:2011}. These are defined as
\begin{equation}
\mathcal{W}_{\text{c}}^{\text{int},\nu}[n_{\text{Dimer}},n_{\text{Atom
1}},n_{\text{Atom 2}}] = \mathcal{W}^\nu_{\text{c}}[n_{\text{Dimer}}] -
\mathcal{W}^\nu_{\text{c}}[n_{\text{Atom 1}}]  -
\mathcal{W}^\nu_{\text{c}}[n_{\text{Atom 2}}],  \label{Eq:IntAC}
\end{equation}
where each of the atomic contributions are evaluated in the presence
of the basis functions of the other atom, thereby accounting for the
basis-set superposition error in the calculated difference. These
integrands are presented for both systems in the lower two panels of
Fig.~\ref{fig:vdW}. The trends in both figures are remarkably similar.
The \textit{ab initio} estimates of the interaction ACs show integrands
which become smoothly more negative with increasing interaction
strength. This is similar to the behaviour shown in
Ref.~\cite{Teale:2011} for the (He)$_2$ system at larger internuclear
separations. The behaviour of the BLYP integrand is striking because it
is significantly too negative in the low-$\nu$ regime
before switching to positive values at larger $\nu$ values.
This behaviour is also similar to that observed in
Ref.~\cite{Teale:2011}; that the curves for both systems are so
qualitatively similar reflects the fact that their shape is determined
mainly by their behaviour under uniform coordinate scaling according to
Eq.~(\ref{deltaclambdaexpr_scaled_density_blyp}), rather than the density on which they are evaluated. The interaction ACs for the double hybrids are significantly afflicted by the errors in the LYP contributions. In the first segment of the AC the influence of the density-functional component is clear, particularly for the density scaled variant, even though it is not the dominant contribution to the overall correlation energy AC. It is also notable that all of the double hybrid interaction ACs are significantly too flat in this section. 
In this respect, the double hybrid schemes do not seem to improve BLYP at all,
with respect to the correlation energy, since this component becomes less attractive. Moreover, 
the exchange interaction energy which is repulsive at the BLYP level (+26 $\mu
E_h$ for (He)$_2$ and +57 $\mu E_h$ for He-Ne) becomes attractive at the
double hybrid level (-114 $\mu E_h$ for (He)$_2$ and  -146 $\mu E_h$ for
He-Ne) whilst the corresponding {\it ab initio} values (+74 $\mu E_h$ for
(He)$_2$ and +68 $\mu E_h$ for He-Ne) are clearly
repulsive. Let us keep in mind though that along our approximate double
hybrid AC (i) the reference density is the KS-BLYP one (which is
therefore not affected by the MP2 treatment) and (ii) the density is not
strictly the same
along the AC. Since the interaction energies considered here are very small,
the density constraint might be important and contribute, through the
orbital relaxation, to both exchange and correlation interaction
energies. The density obtained at the double hybrid level should then be
used as reference
for setting up a true AC where the density constraint is indeed fulfilled.
This should
clearly be analyzed further in the future. Nevertheless, when comparing the total
interaction energies, $\lambda_1$-B2-PLYP is less repulsive (+29 $\mu
E_h$ for (He)$_2$ and +13.5 $\mu E_h$ for He-Ne) than BLYP (+136.5 $\mu
E_h$ and +159 $\mu E_h$ for He-Ne).  
The overall result is that the double hybrids based on a linear AC do
too little to improve the description of both (He)$_2$ and He-Ne dimers. 
This rationalizes to some extent the need for empirical dispersion
corrections even when using a functional such as B2-PLYP, see e.g. Ref.~\cite{Schwabe:2007}. 

\subsection{LiH, HF, N$_2$ and H$_2$O molecules}\label{subsec:results:mols}

Finally in this section we examine a number of small molecular systems
close to their equilibrium geometries. The exchange energy contributions
for these systems are shown in Table~\ref{tab:Ex}. For LiH both the HF
and density-functional contributions are close to each other and
reasonable compared to the KS[CCSD(T)] estimates. As a consequence the
average relevant to the double-hybrid approximations is also of similar
accuracy. For the HF, N$_2$ and H$_2$O molecules the HF estimates of the exchange energies are reasonable, whilst the density-functional estimates are too negative in comparison with the KS[CCSD(T)] values. The averaging used in the double hybrid approaches therefore improves the exchange energy estimate relative to the pure density-functional approach. 

\begin{figure}
\centering
{\includegraphics[scale=1]{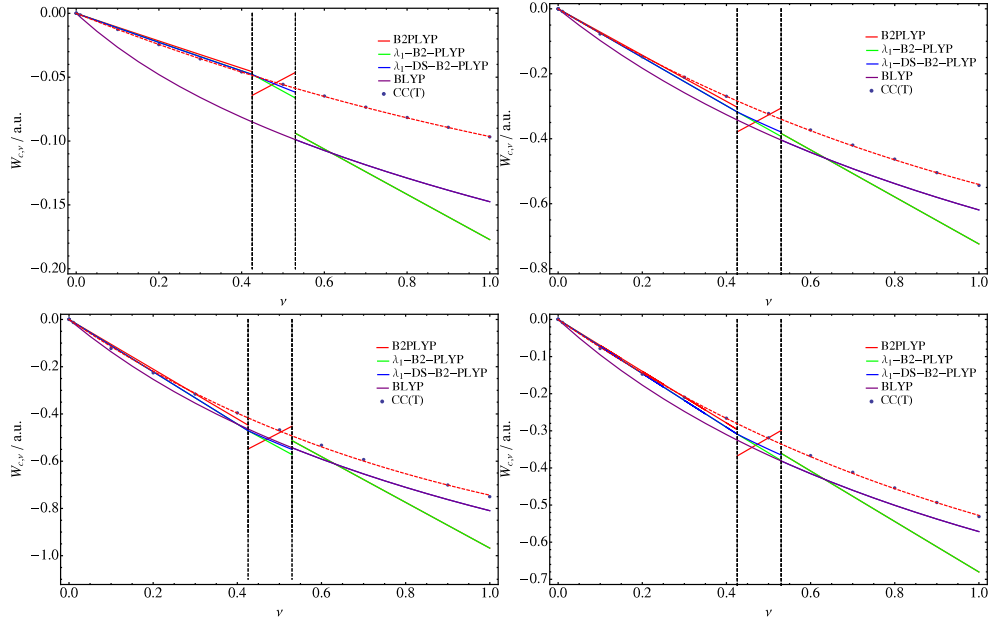}}
\caption{ACs for the molecules LiH (top left panel),
HF (top right panel), N$_2$ (bottom left panel) and H$_2$O (bottom right
panel) calculated using the conventional B2-PLYP functional (solid red
line), the $\lambda_1$-B2-PLYP two parameter double hybrid (green line),
the $\lambda_1$-DS-B2-PLYP functional (blue line) (which includes coordinate scaling contributions) and the standard BLYP functional (purple line). Also included is an accurate \textit{ab initio} estimate of the AC calculated at the CCSD(T) level (blue points). The function in Eq.~(\ref{WACCI}) has been fitted to this data (red dashed line) and the coefficients for this form are reported in Table~\ref{tab:fitcoeffs}.}\label{fig:mol}
\end{figure}

The correlation AC integrands for LiH, HF, N$_2$ and H$_2$O are shown in the four panels of Fig.~\ref{fig:mol}. The plots for HF, N$_2$ and H$_2$O are qualitatively similar to those for the van der Waals dimers and the H$_2$ molecule at $R=1.4$ a.u. The \textit{ab initio} estimates of the correlation AC integrand are relatively subtly curved for each of these species, reflecting the shape expected for systems dominated by dynamical correlation close to their equilibrium geometries. The double-hybrid model correlation ACs for these systems follow similar trends to those discussed in Sections~\ref{subsec:results:H2} and~\ref{subsec:results:vdW}. It is perhaps noteworthy that the initial part of the ACs up to $\lambda_1$ for these systems is reasonably well described by PT2 theory on the $\lambda_1$-interacting system. 

The LiH molecule is more challenging for the DFT and double-hybrid
models. The \textit{ab initio} estimate is typical of a dynamically
correlated system close to its equilibrium geometry. However, the BLYP
functional gives a much too negative correlation integrand with a much
too steep initial slope. This error is typical for many density
functional approximations, which struggle to describe diatomic molecules composed
of group 1 elements. These species typically have rather long equilibrium bond lengths compared to diatomic molecules composed of main group elements. In the first part of the model double-hybrid ACs it is clear that the $\lambda_1$-interacting PT2 estimate is reasonable and is close to the \textit{ab initio} estimate. In the intermediate region as more DFT contributions are included the correlation ACs for the $\lambda_1$ variants begin to slightly overestimate the correlation integrand. In the third section the double hybrid models inherit the large errors present in the density functional description of the correlation integrand for this species. The striking difference in behaviour between the correlation AC models for the LiH molecule and the other species considered here highlights the difficulty in developing a truly transferable double-hybrid approach suitable for a wide range of systems, even when close to their respective equilibrium geometries. However, for a large number of species close to their equilibrium structures double-hybrid approaches are expected to be accurate. In future we hope that the AC analysis presented here can be extended to a much larger set of molecules and used to effectively evaluate double hybrid approaches based on a variety of wave-function and density-functional components. This approach could be effective in identifying and avoiding models which rely on large error cancellations and provide more stringent tests of the models than just post construction benchmarking against experimental data.

\section{Conclusions}\label{sec:conclusions}

In this work we have explicitly derived the AC 
integrands underlying double-hybrid approaches. The integrands have been calculated for the conventional
B2-PLYP scheme and its so-called $\lambda_1$ variant which was
obtained from second-order density-functional perturbation
theory. These integrands were then compared graphically 
with benchmark \textit{ab initio} estimates to assess their accuracy and a number of interesting features have been highlighted.
The approximate one- and two-parameter double hybrid ACs were divided into
three segments, the first up to $\lambda_1$ being dominated by wave
function theory contributions at the PT2 level, the second between
$\lambda_1$ and $\lambda_2$ involving both wave function and
density-functional contributions and the final section beyond
$\lambda_2$ involving purely density-functional contributions. 


Within each section of the approximate ACs the impacts of the approximations utilized in their derivations have been highlighted. In the first section the use of PT2 theory based on fixed orbitals gives a linear approximation to the AC and its slope can be understood from the nature of the orbitals determined for the $\lambda_1$ interacting system. In the intermediate region the behaviour of recently developed two-parameter forms sharply contrasts that of empirical forms, the latter giving ACs with positive slopes. In the final section, which is determined by density-functional contributions, the neglect of uniform coordinate scaling effects has been highlighted and leads to a linear approximation of the correlation AC. Inclusion of these effects can restore some curvature in the integrand, although the outright accuracy still depends heavily on the density-functional form employed.  

The most striking feature of the approximate double hybrid correlation integrands is the presence of (derivative) discontinuities at the connecting $\lambda_1$ and $\lambda_2$ interaction strengths. Whilst these discontinuities do not affect the calculation of correlation energies (only right continuity is required~\cite{Teale:2010}), they may have implications for the determination and uniqueness of the multiplicative potentials associated with keeping the density constant along the AC. In future the design of double-hybrid models which avoid these features in their AC integrands should be considered.

In the context of the van der Waals dimers the difficulties in
constructing a double-hybrid approach based on the linear AC have been
rationalized in terms of the interaction ACs. Here the failure of
$\lambda_1$-B2-PLYP and B2-PLYP approaches to account for longer-ranged
interactions is evident and results in a significant underestimation of the correlation interaction
energy. Recently, it has been shown that range-dependent generalized ACs can leverage physical insight about the range of interactions in this context to more effectively divide labour between the density-functional and wave function components~\cite{Teale:2011}. We also note that the techniques employed in this work may be directly carried over to the analysis of range-separated double-hybrid methods by choosing an alternative non-linear AC integration path. 


Finally, we remark that the integrands presented here highlight that for
more successful double-hybrid approaches it is essential to seek both
improved wave function and density-functional components. In
Section~\ref{subsec:results:H2} we highlighted the RPA based methods as
one possible route to include terms of higher-order in the interaction
strength. Clearly it is a challenging task to introduce such
higher-order contributions without incurring significantly increased
computational cost. An equally challenging task is the construction of
density functional components more compatible with these orbital based
methodologies. It remains to be investigated if forms based on
correlation functionals other than LYP can be more effective in this
sense. The techniques used here could also be extended to further segment the AC in order to
rationalize the behaviour of double hybrids with three or more parameters (see, for example,
Refs.~\cite{Xin_XYG3_2009,Xin_XYGJ-OS_2011,Xin_Adamo_xDH-PBE0}). The \textit{ab initio} estimates of the AC integrands may provide useful guidance in the development of these approaches and we expect that the type of analysis outlined here can play a central role in the future development of more robust double-hybrid approximations.  

\section*{Acknowledgments}
The authors are pleased to dedicate this work to Trygve Helgaker on the occasion of his 60th birthday. 
A.M.T. is grateful for support from the Royal Society University Research
Fellowship scheme, the Norwegian Research Council Grant No. 179568/V30
for the Centre for Theoretical and Computational Chemistry and the
European Union's Seventh Framework Programme (FP7/2007-2013)/ERC Grant
agreement No. 267683. E.F. thanks ANR (DYQUMA project) as well as Pr.
Xin Xu for fruitful discussions.

\appendix

\section{Integration of the exact segmented integrand}\label{Asec:IntegSegAC}

Let us introduce the $\lambda$-dependent decomposition of the exact ground-state energy 
\begin{eqnarray}\label{gsener_lambda-wf}
E^\lambda&=&
\displaystyle
\int^1_\lambda
\frac{\mathrm{d}\mathcal{E}^\nu}{\mathrm{d}\nu}\;\mathrm{d}\nu+\mathcal{E}^\lambda\nonumber\\
&=&\langle \Psi^{\lambda}\vert
\hat{T}+\lambda\hat{W}_{\rm ee}+\hat{V}\vert\Psi^{\lambda}\rangle
+\overline{E}^{\lambda}_{\rm Hxc}[n_{\Psi^{\lambda}}]
,
\end{eqnarray}
where the {complement} $\lambda$-interacting Hxc density-functional
energy equals
\begin{eqnarray}\label{complambdaintHxc_integration}
\overline{E}^{\lambda}_{\rm Hxc}[n]&=&\displaystyle
\int^1_\lambda
\langle \Psi^{\nu}\vert
\hat{W}_{\rm ee}\vert\Psi^{\nu}\rangle
\;\mathrm{d}\nu
\nonumber\\
&=& 
\int^1_0
\langle \Psi^{\nu}\vert
\hat{W}_{\rm ee}\vert\Psi^{\nu}\rangle
\;\mathrm{d}\nu
-\int_0^\lambda
\langle \Psi^{\nu}\vert
\hat{W}_{\rm ee}\vert\Psi^{\nu}\rangle
\;\mathrm{d}\nu,
\end{eqnarray}
this leads, according to Eqs.~(\ref{Hxcexactintegrand}) and
(\ref{lambdaintC_integration}), to the expression given in
Eq.~(\ref{Hxcm1decomp}). In the exact theory, the energy
$E^\lambda$ does not
depend on $\lambda$. Still, it is convenient to derive its derivative with respect to
$\lambda$. It is obtained from the variational
expression of the energy 
\begin{eqnarray}\label{energyminpsi}\begin{array} {l}
{\displaystyle
E^{\lambda} = \underset{\Psi}{\rm min}\left\{ \langle \Psi\vert
\hat{T}+\lambda\hat{W}_{\rm ee}+\hat{V}\vert\Psi\rangle
+\overline{E}^{\lambda}_{\rm Hxc}[n_{\Psi}]\right\}  
},
\end{array}
\end{eqnarray}
and the Hellmann-Feynman theorem which leads to 
\begin{eqnarray}\label{hellmann_feynman_true_ener_exact}
{\displaystyle
\frac{\mathrm{d}E^{\lambda}}{\mathrm{d}\lambda}}&=& 
\langle \Psi^{\lambda}\vert
\hat{W}_{\rm ee}\vert\Psi^{\lambda}\rangle
+{\displaystyle\frac{\partial \overline{E}^{\lambda}_{\rm Hxc}}{\partial
\lambda}[n_{\Psi^{\lambda}}]
}
\nonumber\\
&=&
\langle \Psi^{\lambda}\vert
\hat{W}_{\rm ee}\vert\Psi^{\lambda}\rangle
-E_{\rm Hx}[n_{\Psi^{\lambda}}]-\Delta^{\lambda}_{\rm
c}[n_{\Psi^{\lambda}}]
.
\end{eqnarray}
When integrating the segmented exchange--correlation integrand in
Eqs.~(\ref{segmented_EXXintegrand}) and
(\ref{exact_integrand_lambda1-lambda2}), we therefore obtain from
Eqs.~(\ref{DS1H_reduction_lambdaeqzero}),
(\ref{HxcexactintegrandHxcdecomp}), and
(\ref{hellmann_feynman_true_ener_exact}), 
\begin{eqnarray}\label{energyfromintsegments}
E&=&E^0+
\displaystyle
\int^{\lambda_1}_0
\frac{\mathrm{d}{E}^\nu}{\mathrm{d}\nu}\;\mathrm{d}\nu
\nonumber\\
&&+\int_{\lambda_1}^{\lambda_2}
\Bigg(
\langle \Psi^{\lambda_1}\vert
\hat{W}_{\rm ee}\vert\Psi^{\lambda_1}\rangle
-E_{\rm Hx}[n_{\Psi^{\lambda_1}}] 
-\Delta^{\lambda_1}_{\rm
c}[n_{\Psi^{\lambda_1}}]
\Bigg)\;\mathrm{d}\nu,
\end{eqnarray}
which leads to Eq.~(\ref{gsener_lambda12-wf}).

\section{Perturbation expansion of the exact segmented integrand}\label{Asec:PTexpSegAC}
According to Eqs.~(\ref{complambda12Hxcfun}) and (\ref{ds2dhenerexp}), in the $\alpha=1$ limit, the energy ${E}^{\alpha,\nu,\nu}$ reduces through second order
to 
\begin{eqnarray}\label{Ept2nunu}
E^{[2]\nu}
&=&  
\langle \Phi^{\nu}\vert
\hat{T}+\nu\hat{W}_{\rm ee}+\hat{V}\vert\Phi^{\nu}\rangle
+\overline{E}^{\nu}_{\rm Hxc}[n_{\Phi^{\nu}}]
+\nu^2
E^{(2)\nu}_{\mbox{\tiny
MP}}.
\end{eqnarray}
From Eq.~(\ref{calc_philambda1}) and the Hellmann-Feynman theorem, we obtain the first-order
derivative expression 
\begin{eqnarray}\label{dEpt2nunudnu}
\frac{\mathrm{d}E^{[2]\nu}}{\mathrm{d}\nu}
&=&  
\langle \Phi^{\nu}\vert
\hat{W}_{\rm ee}\vert\Phi^{\nu}\rangle
-{E}_{\rm Hx}[n_{\Phi^{\nu}}]
-{\Delta}^\nu_{\rm c}[n_{\Phi^{\nu}}]
+\frac{\mathrm{d}}{\mathrm{d}{\nu}}\Big(\nu^2
E^{(2)\nu}_{\mbox{\tiny
MP}}\Big).
\end{eqnarray}
In addition, in the $\alpha=1$ limit, the first-order derivative in
the third term on the right hand side of
Eq.~(\ref{segments_auxener_pt}) reduces through second order to
\begin{eqnarray}\label{dEpt2lam1nudnu}
\frac{\mathrm{d}E^{[2]\lambda_1,\nu}}{\mathrm{d}\nu}
&=&  
\langle \Phi^{\lambda_1}\vert
\hat{W}_{\rm ee}\vert\Phi^{\lambda_1}\rangle
-{E}_{\rm Hx}[n_{\Phi^{\lambda_1}}]
-{\Delta}^{\lambda_1}_{\rm c}[n_{\Phi^{\lambda_1}}]
+2\lambda_1
E^{(2)\lambda_1}_{\mbox{\tiny
MP}}
\nonumber\\
&&
\displaystyle
-
\int \mathrm{d}\mathbf{r}\;
\Bigg(
\frac{\delta{E}_{\rm Hx}}{\delta
n(\mathbf{r})}[n_{\Phi^{\lambda_1}}]
+
\frac{\delta{\Delta}^{\lambda_1}_{\rm c}}{\delta
n(\mathbf{r})}[n_{\Phi^{\lambda_1}}]
\Bigg)
\delta
n^{(2)\lambda_1}(\mathbf{r}).
\end{eqnarray}
Note that, in order to compute the MP2 term in Eq.~(\ref{dEpt2nunudnu}), one would in
principle need the response of the orbitals and their energies to 
the variations of $\nu$. Instead, we replace the $\nu-$dependent MP2
correlation energy by its value at $\nu=\lambda_1$ which, after
integration over [0,$\lambda_1$], gives the same result:
\begin{eqnarray}\label{integrationmp2term}
\displaystyle
\int^{\lambda_1}_0
\frac{\mathrm{d}}{\mathrm{d}\nu}\Bigg(\nu^2E^{(2)\nu}_{\mbox{\tiny
MP}}\Bigg)\;\mathrm{d}\nu
&=&
\int^{\lambda_1}_0
2\nu E^{(2)\lambda_1}_{\mbox{\tiny MP}}\;d\nu.
\end{eqnarray}
Combining all equations with Eq.~(\ref{segmented_EXXintegrand}) leads to the second-order expansion of the
correlation integrand given in Eq.~(\ref{ds2dh_integrand}).

\section{Correlation integrand associated to B2-PLYP}\label{Asec:B2PLYPInt}

Since $\overline{E}^{\nu,\nu^2}=\tilde{E}^{\nu,\nu}$, the second and
third term
in the right-hand side of
Eq.~(\ref{convb2plyp_segmentation}) are derived exactly like in
the $\lambda_1$-B2-PLYP scheme, using the Hellmann-Feynman theorem.
Similarly, we obtain
\begin{eqnarray}\label{Conv2H_Hell-Feyn3}
\displaystyle
\frac{\mathrm{d}}{\mathrm{d}\nu}
\overline{E}^{a_{\rm x},a^2_{\rm x}-(a_{\rm x}-\nu)^2}
&=&
-
2(a_{\rm x}-\nu)
{E}^{\mbox{\tiny LYP}}_{\rm
c}[n_{\overline{\Phi}^{a_{\rm x},a^2_{\rm x}-(a_{\rm x}-\nu)^2}}]
\nonumber\\
&&+
\frac{\mathrm{d}}{\mathrm{d}\nu}
\Bigg(
\big(a^2_{\rm x}-(a_{\rm x}-\nu)^2\big)
\overline{E}^{(2)a_{\rm x},a^2_{\rm x}-(a_{\rm x}-\nu)^2}_{\mbox{\tiny
MP}}
\Bigg).
\end{eqnarray}
In order to avoid the calculation of the orbital response to
variations of $\nu$, we gather all MP2 contributions as follows,
\begin{eqnarray}\label{convb2plyp_in_mp2terms}
\displaystyle
&&\int^{a_{\rm x}-\sqrt{a^2_{\rm x}-a_{\rm c}}}_0
\;
\frac{\mathrm{d}}{\mathrm{d}\nu}
\Big(\nu^2\overline{E}_{\mbox{\tiny MP}}^{\nu,\nu^2}
\Big)
\mathrm{d}\nu
\nonumber\\
&+&
\displaystyle
\int^{a_{\rm x}}_{a_{\rm x}-\sqrt{a^2_{\rm x}-a_{\rm c}}}
\;
\frac{\mathrm{d}}{\mathrm{d}\nu}\Bigg(
\nu^2\overline{E}_{\mbox{\tiny MP}}^{\nu,\nu^2}
-
\big(a^2_{\rm x}-(a_{\rm x}-\nu)^2\big)
\overline{E}^{(2)a_{\rm x},a^2_{\rm x}-(a_{\rm x}-\nu)^2}_{\mbox{\tiny
MP}}
\Bigg)
\;\mathrm{d}\nu
\nonumber\\
&=&
a_{\rm c}\overline{E}_{\mbox{\tiny MP}}^{a_{\rm x},a_{\rm
c}}\nonumber\\
&=&
\displaystyle
\int^{a_{\rm x}-\sqrt{a^2_{\rm x}-a_{\rm c}}}_0
\;
2\nu\overline{E}_{\mbox{\tiny MP}}^{a_{\rm x},a_{\rm c}}
\mathrm{d}\nu
+
\displaystyle
\int^{a_{\rm x}}_{a_{\rm x}-\sqrt{a^2_{\rm x}-a_{\rm c}}}
\;
2\big(
a_{\rm x}-\sqrt{a^2_{\rm x}-a_{\rm c}}\big)
\overline{E}_{\mbox{\tiny MP}}^{a_{\rm x},a_{\rm c}}
\;\mathrm{d}\nu,
\end{eqnarray}
which finally leads to the correlation integrand expression in
Eq.~(\ref{convb2plyp_Cintegrand}).


\bibliographystyle{tMPH}


\newcommand{\Aa}[0]{Aa}


\label{lastpage}

\end{document}